\documentclass{article} 
\usepackage{iclr2026_conference,times}

\usepackage{amsmath,amsfonts,bm}

\def\eqref#1{equation~\ref{#1}}

\def\1{\bm{1}}

\DeclareMathAlphabet{\mathsfit}{\encodingdefault}{\sfdefault}{m}{sl}
\SetMathAlphabet{\mathsfit}{bold}{\encodingdefault}{\sfdefault}{bx}{n}

\usepackage{hyperref}
\usepackage{url}

\usepackage{enumitem} 
\usepackage{xcolor}   
\usepackage[dvipsnames]{xcolor}
\usepackage{booktabs}
\usepackage{amsmath}
\usepackage{makecell}
\usepackage{microtype}
\usepackage{wrapfig}

\usepackage{multirow}
\usepackage{graphicx} 
\usepackage[table]{xcolor}
\usepackage{tcolorbox}

\newcommand{\grayrule}{\arrayrulecolor{black!30}\midrule[0.1pt]\arrayrulecolor{black}}

\usepackage{CJKutf8}

\usepackage{float}

\title{Agentic Cognitive Profiling: Realigning Automated Alzheimer’s Disease Detection with Clinical Construct Validity}

\author{
\hspace{-0.2em}\parbox{\textwidth}{
Jiawen Kang,
Kun Li,
Dongrui Han,
Jinchao Li,
Junan Li,
Lingwei Meng,
\\
Xixin Wu, Helen Meng}
\\
The Chinese University of Hong Kong\\
\texttt{jwkang@se.cuhk.edu.hk}
}

\iclrfinalcopy 
\begin{document}

\maketitle


\begin{abstract}

Automated Alzheimer's Disease (AD) screening has predominantly followed the inductive paradigm of pattern recognition, which directly maps the input signal to the outcome label. This paradigm sacrifices construct validity of clinical protocol for statistical shortcuts. 
This paper proposes Agentic Cognitive Profiling (ACP), an agentic framework that realigns automated screening with clinical protocol logic across multiple cognitive domains. Rather than learning opaque mappings from transcripts to labels, the framework decomposes standardized assessments into atomic cognitive tasks and orchestrates specialized LLM agents to extract verifiable scoring primitives. 
Central to our design is decoupling semantic understanding from measurement by delegating all quantification to deterministic function calling, thereby mitigating hallucination and restoring construct validity. 
Unlike popular datasets that typically comprise around a hundred participants under a single task, we evaluate on a clinically-annotated corpus of 402 participants across eight structured cognitive tasks spanning multiple cognitive domains.
The framework achieves 90.5\% score match rate in task examination and 85.3\% accuracy in AD prediction, surpassing popular baselines while generating interpretable cognitive profiles grounded in behavioral evidence. This work demonstrates that construct validity and predictive performance need not be traded off, charting a path toward AD screening systems that explain rather than merely predict.

\end{abstract}
\section{Introduction}

Alzheimer’s Disease (AD) is a progressive neurodegenerative disorder characterized by the deterioration of specific cognitive faculties, such as memory and executive function. Since neuropathological changes are often clinically silent in early stages, clinicians rely on standardized cognitive tests—such as the Montreal Cognitive Assessment (MoCA) ~\citep{hobson2015montreal}—as causal probes to detect underlying deficits~\citep{harvey2012clinical}. 
Unlike passive observation, these tests are engineered protocols designed to isolate and quantify distinct cognitive domains. For instance, list learning tasks explicitly stress memory encoding and retrieval~\citep{delis2000california}, while visual naming tasks probe the integrity of semantic knowledge and visual perception~\citep{giles1996performance}. Consequently, this paradigm of structured cognitive profiling yields verifiable diagnostic evidence, establishing a reliable cornerstone for clinical decision-making.

In contrast to this \textit{deductive} clinical logic, the field of automated AD screening has predominantly followed the paradigm of \textit{inductive} pattern recognition.
Early approaches relied on handcrafted features (e.g., lexical diversity, syntactic complexity)~\citep{fraser2016linguistic, weiner2019speech}, while recent works leverage the Pre-trained Language Models (PLMs)~\citep{balagopalan2020bert, yuan2020disfluencies} to map input transcripts directly to diagnostic labels.
Although PLM-based methods have significantly improved benchmark performance (e.g., on ADReSS~\citep{adress}), they fundamentally treat AD detection as a \textit{"de-contextualized"} classification task, which identifies arbitrarily possible statistical correlations between featurized inputs and diagnostic labels.
These "black-box" models face inherent reliability challenges.
Notable studies have demonstrated the "Clever Hans" effect~\citep{liu2024clever, sahidullah2025shortcut, kang2024within}, where models could predict correct labels based on non-pathological shortcuts.
Besides, the field observes inconsistent findings regarding feature robustness on speech recognition errors~\citep{kang2024not, li2024useful}.
Fundamentally, these issues are stem from their intrinsic paradigm: \textit{pattern recognition models predict outcomes rather than extract causal evidence, as in clinical protocol.}


From the lens of construct validity~\citep{strauss2009construct}, the divergence between these two paradigms reveals a fundamental measurement gap~\citep{van2024undesirable}.
The clinical protocol largely operationalizes the target neurocognitive constructs: performance decrements are directly attributable to specific failures in cognitive domains (e.g., memory deficits).
Conversely, predominant data-driven approaches often sacrifice construct validity for predictive validity, thereby encoding more construct-irrelevant variance—confounding proxy features (e.g., dialect, or acoustic shortcuts) with genuine pathological signals.



This paper aims to realign automated AD screening with clinical grounding. 
To this end, we propose \textit{Agentic Cognitive Profiling (ACP)}, an Agentic Cognitive Assessment Framework that shifts from inductive pattern recognition to deductive clinical profiling across multiple cognitive domains.
The key insight is to operationalize clinical scoring logic through a multi-agent workflow, decoupling semantic understanding from measurement by delegating all quantification to deterministic functions, ensuring verifiable and construct-valid scoring.
We make the following contributions:
\begin{itemize}[leftmargin=*, itemsep=2pt, topsep=2pt]
\item \textbf{Multi-faceted Assessment}: 
Existing public benchmarks for automated AD detection (e.g., ADReSS, N=156) are typically built upon a single elicitation task, capturing a limited facet of cognitive manifestation.
We evaluate on a clinically-annotated corpus of 402 participants across eight structured tasks spanning multiple cognitive domains, providing the first empirical evidence for the feasibility and effectiveness of automated cognitive scoring and AD screening under a multi-domain structured assessment setting.

\item \textbf{Framework}: 
We propose an Agentic Cognitive Assessment Framework that, for the first time, realigns automated AD screening with clinical construct validity. Rather than learning opaque mappings from transcripts to labels, the framework decomposes standardized assessments into atomic cognitive tasks and orchestrates specialized agents to extract verifiable scoring primitives, supporting both zero-shot and supervised screening with interpretable cognitive profiles.
\item \textbf{Performance}: The framework achieves 90.5\% score 
match rate in task examination and 85.3\% accuracy in AD screening, 
surpassing PLM-based baselines in both zero-shot and supervised 
settings while maintaining full interpretability.
\end{itemize}

\section{Related Work}
\begin{figure}[tbp]
\begin{center}
\includegraphics[width=0.45\linewidth,scale=1.0]{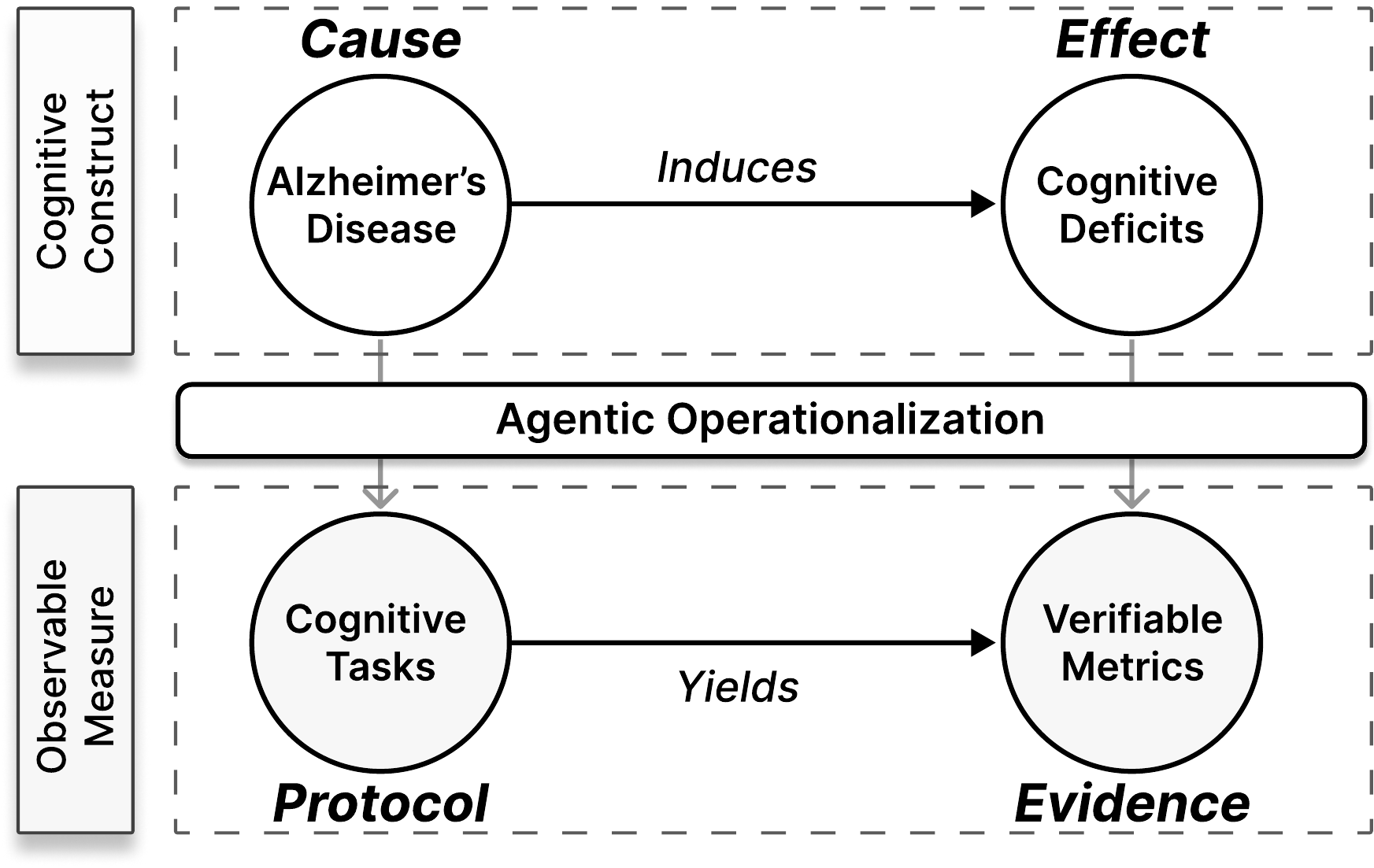}
\end{center}
\vspace{-10pt}
\caption{
The Conceptual Framework. We align AD detection with clinical construct validity by operationalizing the causal chain from Alzheimer’s Disease to Cognitive Deficits into an agentic workflow comprising Cognitive Tasks and Verifiable Metrics.
}
\vspace{-10pt}
\label{fig:highlight}
\end{figure}

\vspace{-5pt}

\subsection{Clinical Cognitive Assessment}
\vspace{-5pt}
Clinical cognitive assessments are standardized protocols designed to isolate and quantify neurocognitive deficits~\citep{harvey2012clinical}.
These instruments generally fall into three categories based on granularity:
\textit{(a) Rapid screening tests}, such as the AD8~\citep{galvin2005ad8}, typically question daily functioning (e.g., troubles with making decisions) and detect initial functional shifts before formal testing;
\textit{(b) Domain-specific tests} target distinct faculties. In language function, the "Cookie Theft" task (from the Boston Diagnostic Aphasia Examination~\citep{goodglass2001bdae}) underpins benchmarks like ADReSS~\citep{adress, adresso}.
In memory function, list-learning tasks ~\citep{delis2000california, au2003verbal} serve as the gold standard for episodic memory;
\textit{(c) Comprehensive screening tests} profile impairment across multiple domains. 
The Mini-Mental Status Examination~\citep{cockrell2002mini} is a widely used baseline, but it suffers from ceiling effects in early-stage pathology. 
Consequently, the Montreal Cognitive Assessment (MoCA)~\citep{hobson2015montreal, yeung2014validation} has emerged as the standard for detecting Mild Cognitive Impairment (MCI) due to its sensitivity to executive dysfunction.
We adopted MoCA and the Hong Kong List Learning Test (HKLLT)~\citep{au2003verbal} as the clinical foundation for this work.

\subsection{Automatic AD detection}
\vspace{-5pt}
Dominant research in natural language-based AD detection has been driven by the exploration of effective features to improve discrimination.
Early efforts utilized handcrafted acoustic and linguistic features~\citep{fraser2016linguistic, weiner2019speech}, such as decreased jitter and lexical diversity.
The advent of pre-trained models has shifted the paradigm toward deep representation learning, achieving state-of-the-art performance across text~\citep{balagopalan2020bert, pipeline, yuan2020disfluencies, yi22}, speech~\citep{haulcy2021classifying, jinchao23, wavbert}, and multi-modal settings~\citep{koo2020exploiting,li2025detecting, syed2021automated} by capturing rich semantic dependencies.

More recently, distinct approaches have explored incorporating certain clinical knowledge as priors in model design.
\cite{li2024devising, park2025reasoning} assess information retrieval in the Cookie Theft picture description task, and \cite{li2025detecting} quantifies image-narrative alignments in image-based storytelling tasks.
These methods provide a degree of  explainability yet lack explicit modeling of clinical decision making, thus limiting both transparency and coverage of cognitive profiling.

Beyond algorithmic advances, a body of research has also validated the deployment practices of automated AD detection systems, including \cite{ding2022talktive, an2025conversational, breithaupt2025designing}. These efforts adopt a human-computer interaction perspective, exploring the use of graphical user interfaces (GUIs) to administer test items to participants. Such work has progressively narrowed the gap between GUI-based interaction and human-administered assessment.

\begin{figure*}[!t]
\begin{center}
\includegraphics[width=1\textwidth]{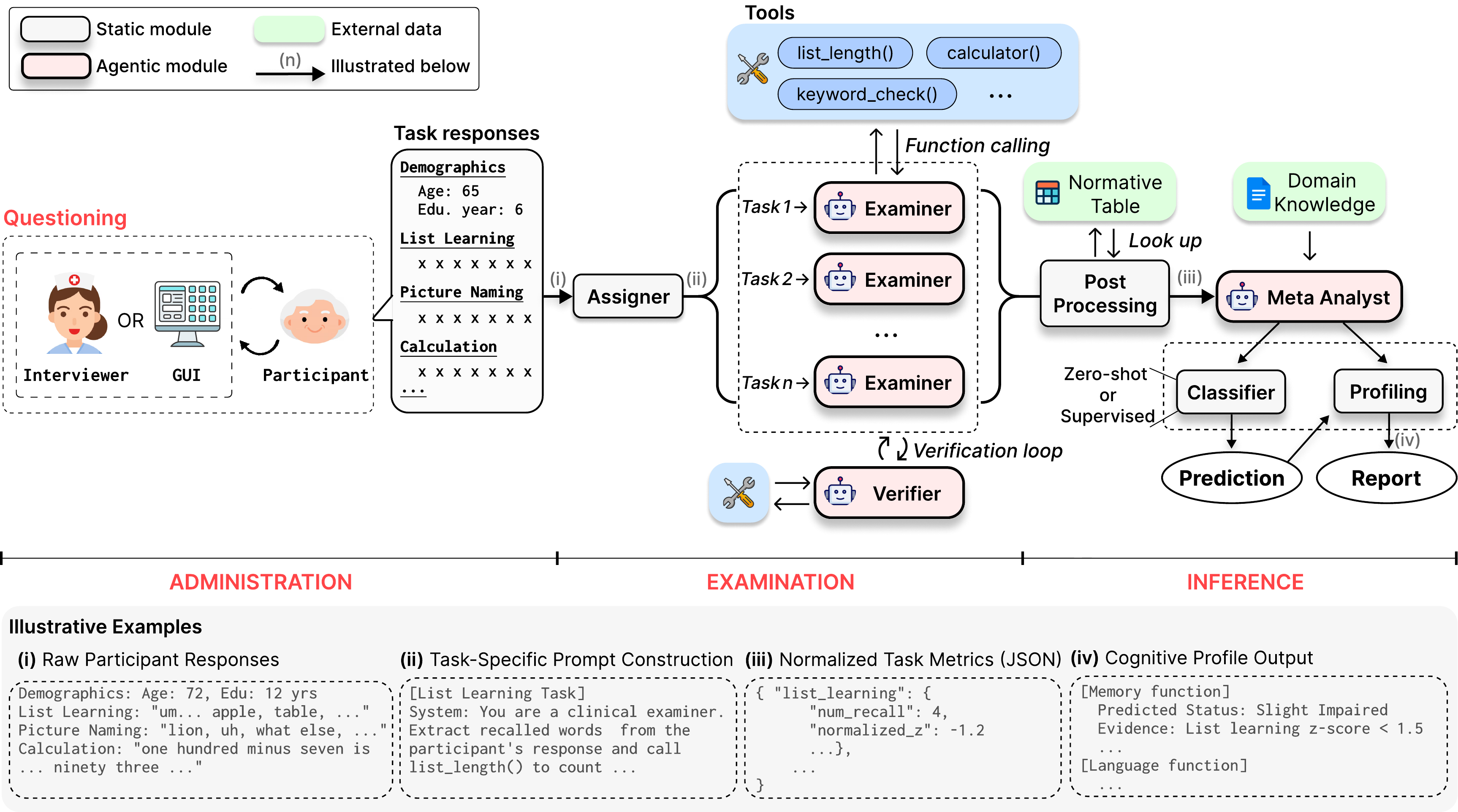}
\end{center}
\vspace{-10pt}
\caption{
Overview of the Agentic Cognitive Assessment Framework. 
The workflow comprises three stages: 
(1) Administration: Collection of standardized task responses; 
(2) Examination: Multi-agent Workflow with Deterministic Function Calling and Verification Loop; 
(3) Inference: Aggregation of verified metrics for classification and explainable reporting.	
}
\vspace{-15pt}
\label{fig:workflow}
\end{figure*}



\subsection{LLM-based Agentic Workflow}
\vspace{-5pt}
Agentic workflows enable LLMs with the capacity to navigate complex problem spaces through iterative cycles of perception, planning, and execution. 
This shift turns opaque generation into transparent, verifiable actions.
Agents have shown great potential in multiple domains, including coding automation \citep{yang2024sweagent}, scientific discovery \citep{m2024augmenting}, and open-world exploration \citep{wang2023voyager}.

In the context of clinical cognitive assessment, very few works have explored this area. 
\citep{bazgir2025agent} employs LLM agents for AD disease management.
\citep{li2025care} and \citep{hou2025adagent} develop agents for data preprocessing and model selection, respectively, while still relying on external neural networks for classification. 



\section{Methodology}
\vspace{-5pt}

\subsection{Task Definition and Cognitive Constructs}
\vspace{-5pt}

We formulate the automatic AD detection process as a cognitive profiling task.
Formally, given a session transcript $X$, our objective is to extract a set of interpretable \textit{Scoring Primitives}, $\mathcal{S} = \{s_1, s_2, ..., s_k\}$, representing the atomic units of clinical evidence (e.g., "successful naming a depicted animal", or "successful recall of `velvet'").
These primitives serve as a grounded representation to derive both a human-readable screening report and a binary detection label.


\begin{table}[t]
\centering
\caption{Overview of Cognitive Tasks and Scoring Primitives.
Scoring primitives define measurable performance units and their maximum scores.
}
\setlength{\tabcolsep}{10pt}
\label{tab:cognitive_tasks}
\setlength{\tabcolsep}{15pt}
\scalebox{0.85}{
\begin{tabular}{ >{\raggedright\arraybackslash}p{3.2cm}  p{5cm} @{} r }
\toprule
\textbf{Task} & \textbf{Scoring Primitives} & \textbf{Score} \\
\midrule
\multicolumn{3}{l}{\textit{MoCA-SL (Assess language, attention, executive functions)}} \\
\midrule
Picture Naming & 
Per-item correctness ($\times$3) & 3 \\

Digit Span & 
Forward/backward correctness & 2 \\

Serial 7 Subtraction & 
\# correct subtractions & 3 \\

Sentence Repetition & 
Per-sentence correctness ($\times$2) & 2 \\

Animal Fluency & 
\# valid animals & 1 \\

Abstraction & 
Per-pair correctness ($\times$2) & 2 \\
 & \multicolumn{1}{r}{\textit{Total:}} & 13 \\
\midrule
\multicolumn{3}{l}{\textit{HKLLT (Assess learning and memory functions)}} \\
\midrule
Trial-4 &
\# recalled words (10-min delay) & 16 \\
Trial-5 &
\# recalled words (30-min delay) & 16 \\
\bottomrule
\end{tabular}}
\vspace{-10pt}
\end{table}

\noindent \textbf{Structured Input}: 
Unlike open-ended conversations where cognitive signals are sparse and entangled, we ground our framework in standardized clinical protocols.
Based on MoCA and HKLLT, we adopt a MoCA-SL (Spoken Language subset of MoCA) and Trial $4/5$ in the HKLLT as our profiling tasks, to ensure construct validity and reliable cognitive profiling.
The corresponding cognitive domains and scoring primitives are summarized in Table~\ref{tab:cognitive_tasks}, and we provide further details in Appendix~\ref{appendix:protocol}.
We assume structured input: participant demographics (age, years of education) paired with verbal responses to each cognitive task.
This input format directly mirrors clinical practice, and it is scalable for large-scale deployment, where a growing body of HCI research~\citep{ding2022talktive, an2025conversational, breithaupt2025designing} has demonstrated that GUI-based interfaces can reliably administer these standardized tests.

\noindent \textbf{Explainable Output.}
Beyond binary prediction ($y \in \{\text{AD}, \text{HC}\}$), our framework generates a structured cognitive profile that details \textit{why} a screening conclusion is reached.
We argue that an interpretable screening system should not be limited to binary prediction.
In clinical practice, assessment of cognitive impairment is never a simple yes-or-no decision, it is equally crucial to understand how the assessment leads to a particular conclusion. 
Such information is essential for patient communication and treatment planning. 
Our framework addresses this by generating cognitive reports that detail task-level evidence alongside clinical interpretations (e.g., failure in delayed recall).

\subsection{Agentic Examination}
\vspace{-5pt}

The Examination stage employs a multi-agent workflow to transform verbal responses into structured scoring primitives (see Figure~\ref{fig:workflow}). We describe each component below.

\subsubsection{Task-Specific Examiner Agents}
\vspace{-5pt}


The Assigner routes each task response to a dedicated Examiner agent equipped with task-specific prompts.
Task prompts are structured, comprising four components:
\textit{(a) Task Introduction}: clinical role and assessment objective; \textit{(b) Guidelines}: processing rules and edge cases;
\textit{(c) Output Format}: response structure; and \textit{(d) Examples}: demonstrations for consistent behavior.
And each Examiner therefore applies task-specific rules to process transcription, outputting task-specific scoring primitives for downstream processing.
Prompt template and examples are provided in Appendix~\ref{appendix:prompts:examiner}.



This modular design enables flexible task-specific processing. 
For instance, the Animal Fluency Examiner requires semantic understanding to parse responses, identify valid animals, and deduplicate lexical variants before counting. 
Others could act as simple bridges—the Sentence Repetition Examiner directly passes the target sentence and transcription to \texttt{keyword\_check()} without intermediate processing. 
While more autonomous planning architectures are feasible, our preliminary experiments indicate that fixed, template-based guidelines yield more \textit{stable} and \textit{input-robust} outputs.



\subsubsection{Deterministic Function Calling}
\vspace{-5pt}

A key design principle is the decoupling of semantic understanding from measurement. 
While LLMs interpret natural language content, such as eliminating disfluencies, understanding dialectal variations, and semantics. 
All quantification is delegated to deterministic functions, mitigating hallucination (i.e., fabrication of unsupported outputs) in numerical outputs.
Specifically, Examiner agents have access to a library of scoring tools. 
Some functions are shared across tasks, while others are task-specific.
For instance, \texttt{keyword\_check()} function checks if certain keywords appear in a list, and could serve both Sentence Repetition and Digit Span tasks.
\texttt{parse\_hkllt()} function extracts HKLLT-specific metrics such as semantic clustering.

\subsubsection{Verification Loop}
\vspace{-5pt}

We observe that Examiner agents exhibit hallucination when processing transcripts—LLMs occasionally fabricate false evidence to justify incorrect scores. 
This issue is particularly pronounced when deploying smaller, locally-hosted models. 
To ensure reliability, each Examiner's result undergoes validation by a \textit{verifier} agent. 
The verifier receives the original transcript alongside the Examiner's output and evaluates correctness.
If discrepancies are detected—such as hallucination or incorrectly parsed responses— verifier agent provides feedback specifying the error, and the Examiner regenerates its output. 
This loop continues for up to 3 iterations, after which the result is accepted regardless of verification status.

\begin{figure}[tbp]
\begin{center}
\includegraphics[width=0.5\linewidth,scale=1.0]{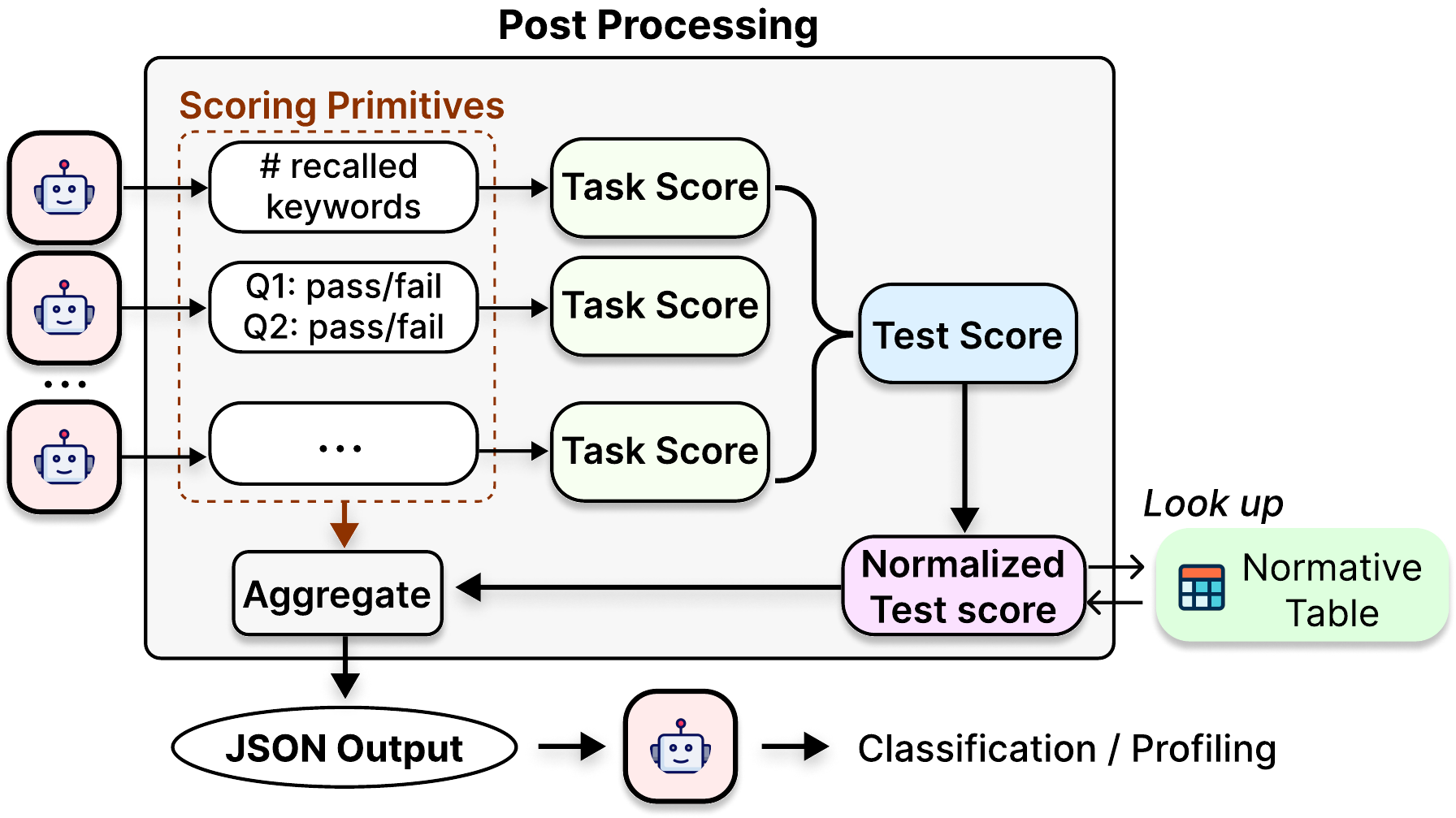}
\end{center}
\vspace{-15pt}
\caption{
Cognitive profile inference pipeline.
Verified scoring primitives from each task are aggregated, normalized against demographic norms, and used for classification and report generation.
}
\vspace{-10pt}
\label{fig:post}
\end{figure}

\subsection{Cognitive Profile Inference}
\vspace{-5pt}

The Inference stage aggregates verified scoring primitives, normalizes them against population norms, and generates both a prediction and an interpretable clinical report (see Figure~\ref{fig:post}).

\subsubsection{Score Normalization}
\vspace{-5pt}

In this stage, scoring primitives from all examiners are collected into a unified JSON structure. 
Task scores are computed by aggregating primitives according to standard clinical protocols, for example, summing per-item correctness flags in Picture Naming task.
These task scores then aggregate into test scores: MoCA-SL (maximum 13 points) and HKLLT trial scores (maximum 16 words each).

To obtain norm-referenced scores, we consult age-education stratified normative tables. 
For HKLLT, published Hong Kong norms are directly available. 
For MoCA-SL, however, existing normative data are based on the full 30-point assessment~\cite{wong2015montreal}. 
We address this by linearly rescaling the full MoCA norms proportionally to the MoCA-SL score range, keeping comparable distributional properties. 
Alternative estimation methods are compared in Appendix~\ref{appendix:norm}.

 \subsubsection{Prediction}
 \vspace{-5pt}

We implement two prediction approaches: zero-shot and supervised. 
The zero-shot method applies established clinical thresholds directly without requiring any training data: a participant is flagged as AD if the MoCA-SL score falls below the 16th percentile, or if either the HKLLT delayed recall score (10-min or 30-min) falls below -1.0 SD. These cutoffs correspond to standard clinical practice for identifying mild cognitive impairment, ensuring full transparency in decision logic.

We also find that behaviorally-grounded scoring primitives serve as effective features for supervised classification. 
Using all primitives as input, we train an SVM classifier that achieves strong performance. Unlike black-box approaches that operate on abstract learned representations, this classifier relies exclusively on interpretable behavioral evidence extracted from task performance. 
 



\subsubsection{Cognitive Profiling}
\vspace{-5pt}

Beyond binary prediction, the Meta Analyst agent explains why a particular outcome is reached, grounding each conclusion in task-level evidence. 
The agent receives the participant's scoring primitives, norm-referenced scores, and a domain knowledge document specifying each task's clinical significance, normal ranges, and interpretation guidelines.

The output comprises two steps: (a) a risk analysis regarding multiple cognitive domains (e.g., memory, executive function), with each domain containing a status indicator, supporting evidence, and clinical interpretation; and (b) a narrative statement summarizing the participant's overall cognitive profile. 
The cognitive profile and final prediction are derived from identical evidence, ensuring that the screening outcome is not merely a label but a transparent, auditable clinical judgment.



\section{Data}
\vspace{-5pt}



To our knowledge, no publicly available dataset provides structured responses from standardized cognitive assessments.
Therefore, we evaluate our framework on an in-house Cantonese speech corpus collected for cognitive screening research. 
The corpus comprises recordings from 1,063 older adults, each participating in a clinician-guided assessment session lasting approximately 1.5 hours. 
Sessions cover multiple standardized cognitive tests, including AD8, MoCA, HKLLT, and the Modified Boston Naming Test (mBNT).


\begin{wrapfigure}{r}{0.5\textwidth} 
\vspace{-10pt}
\includegraphics[width=\linewidth,scale=1.0]{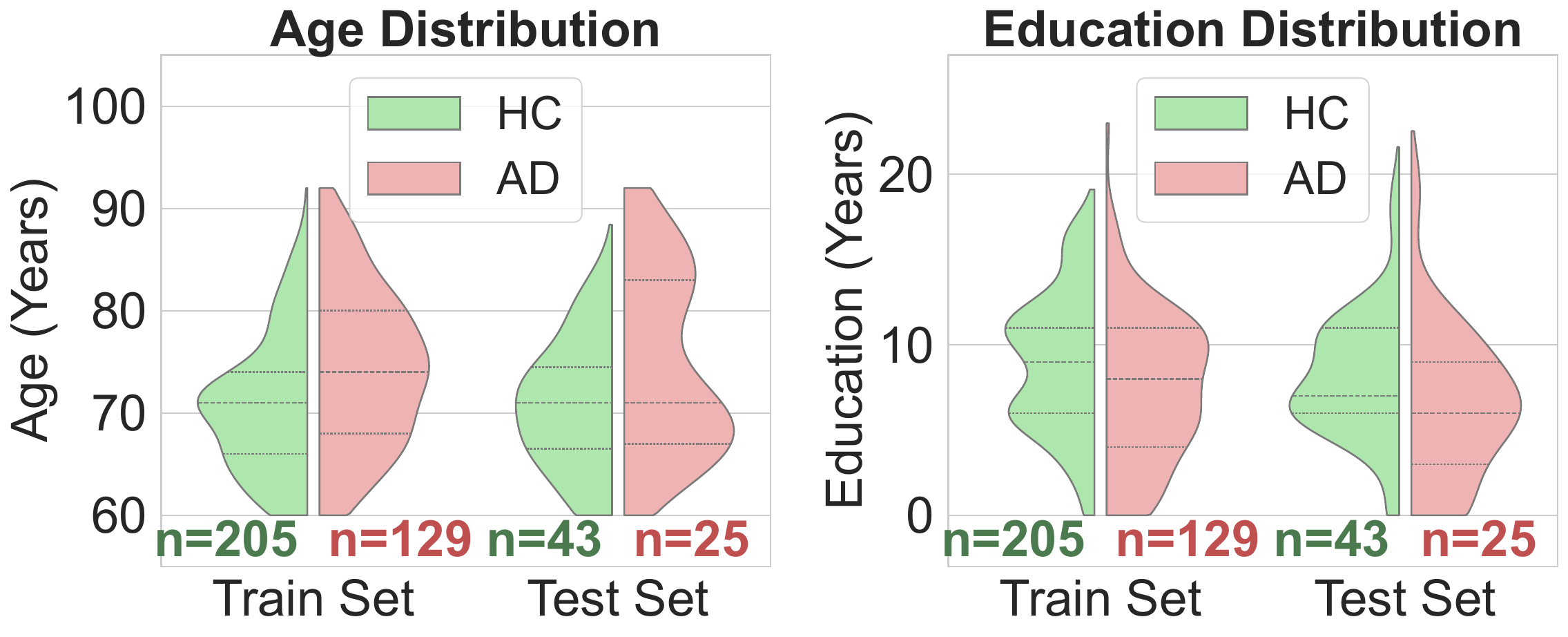}
\vspace{-20pt}
\caption{
Demographic distribution of participants in datasets. Subplots depict age and years of education for Alzheimer’s (AD) and healthy control (HC) groups.
}
\label{fig:demo}
\end{wrapfigure}

\noindent \textbf{Inclusion Criteria.} 
Participants were required to be (1) aged 60 years or above, (2) proficient in spoken Cantonese, and (3) capable of completing all tests with adequate vision and hearing (corrective aids permitted).

\noindent \textbf{Annotation and Labels.} 
Of the full corpus, 402 sessions have been manually transcribed with task-level segmentation. 
Each participant was assigned a binary label---AD-risk (AD) or healthy control (HC)---by professional clinical assessors.
We use this annotated subset for all experiments in this work.

\noindent \textbf{Data Split.} 
We partition the annotated data into training (N=334) and test (N=68) sets. 
The age and education distributions of participants are presented in Figure~\ref{fig:demo}, which shows balanced demographic patterns between the AD and healthy control groups.

\noindent \textbf{Task Coverage.} 
As detailed in Table~\ref{tab:cognitive_tasks}, we extract responses from two test batteries: (1) MoCA-SL, comprising six spoken-language (SL) tasks from the full MoCA protocol, and (2) HKLLT, a representative test assessing learning and memory ability.

\section{Experiments and Results}
\vspace{-5pt}

\subsection{Experimental Setup}
\vspace{-5pt}

We implement all agents using Qwen3-8B deployed locally via vLLM, with temperature set to 0.3 for examiners and 0.1 for verifiers. 
We compare against three baseline categories: (1) traditional approaches using handcrafted linguistic features (13 features); (2) PLM-based methods applying BERT and RoBERTa on concatenated task transcripts; and (3) LLM-CoT, which prompts the same backbone model with concatenated transcripts in a Chain-of-Thought (CoT) manner~\citep{park2025reasoning}.
For task examination, we report Score Match Rate (SMR) and Mean Absolute Error (MAE), where SMR indicates the exact agreement percentage with manual scores; MAE contextualizes error magnitude.
For screening inference, we report Accuracy, F1, Precision, and Recall. 
Supervised classifiers (SVM with RBF kernel, MLP) are trained on extracted scoring primitives. 
All experiments are repeated 5 times, and we report the mean performance.
Implementation details are provided in Appendix~\ref{appendix:implementation}.

\subsection{Task Examination Evaluation}
\label{sec:task_examination_eval}
\vspace{-5pt}

\begin{table}[t]
\centering
\caption{
Task-level examination result.
Score Match Rate (SMR) indicates the exact agreement percentage with manual scores; Mean Absolute Error (MAE) contextualizes error magnitude. Score range is annotated per task. \textit{Func. Call}: Deterministic Function Calling.
}
\label{tab:exam}
\setlength{\tabcolsep}{15pt}
\scalebox{0.85}{
\begin{tabular}{l cc cc cc}
\toprule
 & \multicolumn{2}{c}{\textbf{Full}} & \multicolumn{2}{c}{\textbf{w/o Verifier}} & \multicolumn{2}{c}{\textbf{w/o Func. Call}} \\
\cmidrule(lr){2-3} \cmidrule(lr){4-5} \cmidrule(lr){6-7}
\textbf{Task} & SMR & MAE & SMR & MAE & SMR & MAE \\
\midrule
\multicolumn{7}{l}{\textit{MoCA-SL}} \\
\midrule
Picture Naming {\scriptsize(/3)}   & 97.0\% & 0.03 & 97.0\% & 0.03 & 92.6\% & 0.07 \\
Digit Span {\scriptsize(/2)}       & 98.5\% & 0.01 & 98.5\% & 0.01 & 77.3\% & 0.24 \\
Serial 7 Sub. {\scriptsize(/3)}    & 82.4\% & 0.19 & 70.6\% & 0.32 & 63.2\% & 0.63 \\
Sentence Rep. {\scriptsize(/2)}    & 89.7\% & 0.10 & 89.7\% & 0.10 & 86.8\% & 0.13 \\
Animal Fluency {\scriptsize(/1)}   & 98.5\% & 0.01 & 97.1\% & 0.03 & 98.5\% & 0.02 \\
Abstraction {\scriptsize(/2)}      & 82.3\% & 0.18 & 64.7\% & 0.38 & 67.6\% & 0.32 \\
\midrule
\rowcolor{gray!15}
\textbf{Score-Weighted Avg.}       & 90.5\% & 0.10 & 85.5\% & 0.16 & 79.2\% & 0.27 \\
\midrule
\multicolumn{7}{l}{\textit{HKLLT}} \\
\midrule
Trial-4 {\scriptsize(/16)}         & 94.1\% & 0.07 & 94.1\% & 0.07 & 27.9\% & 2.23 \\
Trial-5 {\scriptsize(/16)}         & 92.6\% & 0.07 & 92.6\% & 0.07 & 23.5\% & 2.32 \\
\bottomrule
\end{tabular}}
\vspace{-10pt}
\end{table}
\noindent \textbf{Examination Results.}
Table~\ref{tab:exam}  presents per-task SMR and MAE. 
Low-inference tasks, such as Picture Naming, Digit Span, and Animal Fluency, achieve near-ceiling accuracy ($>97\%$) as they involve straightforward extraction aligned with clinical scoring rules. 
High-inference task Serial 7 Subtraction and Abstraction show slightly lower but robust performance ($82\%$), as these tasks require nuanced semantic judgment. 
HKLLT delayed recall tasks also achieve high accuracy ($>92\%$). 
Note that Sentence Repetition exhibits lower SMR ($89\%$) due to pronunciation-level ambiguities where clinical assessors and annotators often disagree; we assign lower weight to subsequent analyses. 
Overall, the framework achieves $90.5\%$ weighted-average SMR with 0.10 MAE, demonstrating reliable examination across diverse cognitive tasks.
We leave more results and discussions in Appendix~\ref{appendix:task_eval_details}.


\begin{table}[t]
\centering
\caption{
Comparison of Alzheimer's disease screening performance (\%) across baseline and proposed systems.
Results exceeding 80\% are highlighted in bold.
}
\label{tab:infer}
\setlength{\tabcolsep}{15pt}
\scalebox{0.85}{
\begin{tabular}{lccccc}
\toprule
\textbf{System} & \textbf{Classifier} & \textbf{Accuracy} & \textbf{F1} & \textbf{Precision} & \textbf{Recall} \\
\midrule
\multicolumn{6}{l}{\textit{Baseline}} \\
\midrule
Handcrafted & MLP\,/\,SVM & 68.2\,/\,70.6 & 55.7\,/\,67.8 & 58.3\,/\,56.8 & 54.4\,/\,\textbf{84.0} \\
BERT        & MLP\,/\,SVM & 73.7\,/\,79.4 & 59.5\,/\,72.0 & 68.3\,/\,72.0 & 52.8\,/\,72.0 \\
RoBERTa     & MLP\,/\,SVM & 70.4\,/\,75.0 & 53.9\,/\,66.7 & 63.1\,/\,65.4 & 47.2\,/\,68.0 \\
LLM-CoT     & -            & 70.6          & 66.7          & 57.1          & \textbf{80.0} \\
\midrule
\multicolumn{6}{l}{\textit{Proposed}} \\
\midrule
\rowcolor{gray!15}
\textbf{Zero-shot}          & - & \textbf{82.4} & \textbf{80.2} & \textbf{81.4} & \textbf{80.2} \\
\quad w/o Verifier  & - & 79.4 & 77.4 & 78.1 & 77.5 \\
\quad w/o Function  & - & 69.1 & 63.3 & 66.8 & 63.0 \\
\grayrule
\rowcolor{gray!15}
\textbf{Supervised}          & MLP\,/\,SVM & \textbf{81.5}\,/\,\textbf{85.3} & 66.9\,/\,78.3 & \textbf{95.7}\,/\,\textbf{85.7} & 52.0\,/\,72.0 \\
\quad w/o Verifier  & MLP\,/\,SVM & 73.2\,/\,77.9 & 44.5\,/\,61.5 & \textbf{94.4}\,/\,\textbf{85.7} & 29.6\,/\,48.0 \\
\quad w/o Function  & MLP\,/\,SVM & 74.1\,/\,\textbf{82.4} & 49.8\,/\,76.0 & \textbf{86.6}\,/\,76.0 & 35.2\,/\,76.0 \\
\bottomrule
\end{tabular}}
\vspace{-10pt}
\end{table}

\noindent \textbf{Ablation Study.}
Table~\ref{tab:exam} also quantifies the contribution of the verification loop and deterministic function calling. 
For low-inference tasks, removing either component yields negligible changes, as performance already approaches the ceiling. 
However, on high-inference tasks, the Verifier proves essential: its removal causes substantial drops in Abstraction and Serial 7 Subtraction, indicating it corrects hallucinated evidence in nuanced reasoning. 
Removing function calling causes more severe degradation on counting-intensive tasks (e.g., HKLLT Trial-4), confirming that LLM requires deterministic computation for reliable measurement.
Together, these results demonstrate that both components are critical for challenging tasks.

\noindent \textbf{Case Study.} 
We conduct error analysis to investigate failure modes. 
Details can be found in Appendix~\ref{appendix:case}.
For low-inference tasks, errors primarily stem from model hallucinations that the verifier fails to handle.
As we implemented an 8B model, we expect larger models might be more stable, thus mitigating such issues. 
For high-inference tasks, errors typically occur at boundary cases due to response diversity. 
For instance, in the train-bicycle similarity task, the model accepted "both can move" as correct, while the assessor judged it too generic.

\subsection{Screening Inference Evaluation}
\vspace{-5pt}

\noindent \textbf{Prediction Results.}
Table~\ref{tab:infer} compares AD screening performance across baseline and proposed systems. The proposed agentic framework outperforms prior PLM-based methods in both zero-shot and supervised settings.
Among baselines, BERT achieves the best performance ($79.4\%$ accuracy), consistent with prior findings on PLM-based AD detection. 
LLM-CoT attains $70.6\%$ accuracy in a fully zero-shot manner, demonstrating the advantage of structured clinical inputs for reasoning with domain knowledge—yet its performance remains bounded by the absence of verifiable measurement.
Our proposed framework addresses this limitation through grounded scoring primitives.
The zero-shot approach surpasses the best baseline without any training data, achieving the highest F1 score ($80.2\%$) across all systems. 
The supervised variant further improves accuracy to $85.3\%$, establishing a substantial margin over prior methods. 
We note that SVM consistently outperforms MLP across all configurations, likely due to class imbalance causing naive MLP to exhibit prediction bias toward the majority class. 
Ablation results align with section~\ref{sec:task_examination_eval}: removing the verifier or function calling degrades both zero-shot and supervised performance, confirming their critical role in reliable scoring primitive extraction.

\noindent\textbf{Cognitive Profiling.}
Beyond binary prediction, our framework generates structured cognitive profiles that explain why a screening conclusion is reached. 
Each report is organized by cognitive domain—memory, executive function, attention, and language—with status indicators, supporting evidence, and clinical interpretations grounded in norm-referenced performance. 
This dual-output bridge automated screening with clinical workflow, ensuring that outcomes are transparent and auditable rather than opaque labels.
We showcase several complete example reports in Appendix~\ref{appendix:prompts:analyst}.
 


\subsection{Analysis}
\begin{figure}[tbp]
\begin{center}
\includegraphics[width=0.8\linewidth,scale=1.0,trim=0 0 0 0,clip]{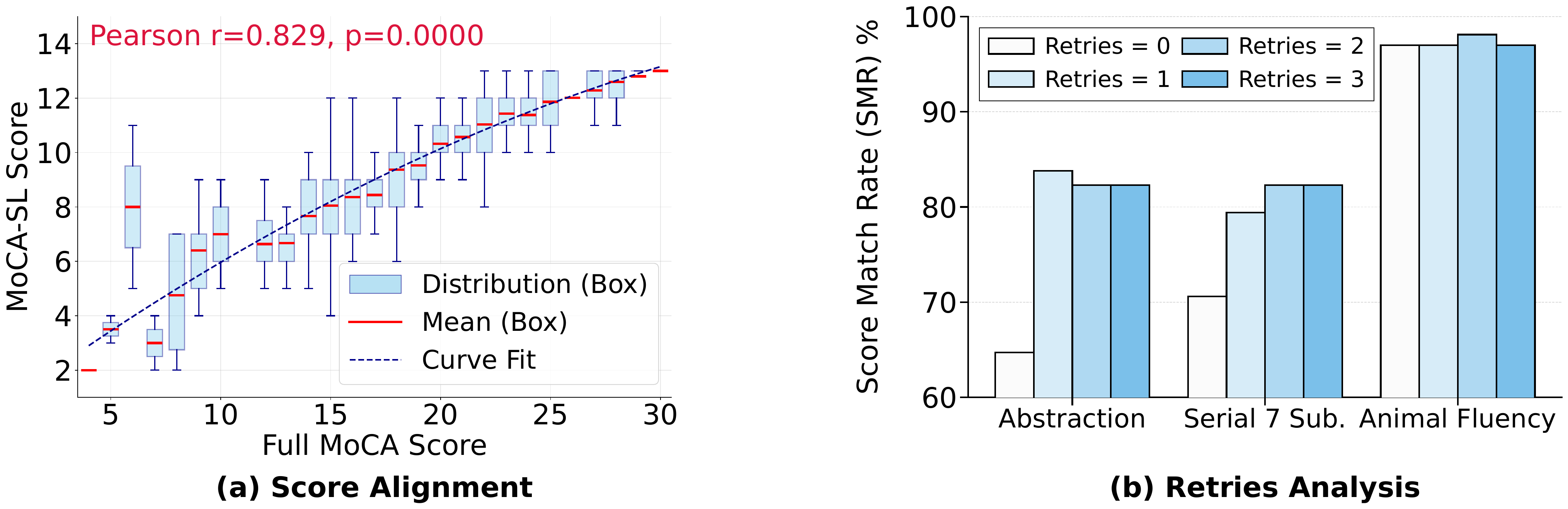}
\end{center}
\vspace{-10pt}
\caption{
(a) Alignment of the full MoCA score and the subset MoCA-SL score; (b) Performance comparison of the Score Match Rate (SMR) with varying maximum retry limits.
}
\vspace{-10pt}
\label{fig:score}
\end{figure}

\vspace{-5pt}
\noindent \textbf{Validate MoCA-SL scores.}
MoCA-SL serves as the core protocol for assessing language, attention, and executive functions, directly influencing both screening outcomes and cognitive profiling. 
As a spoken-language subset, a natural question arises: Does MoCA-SL introduce measurement noise compared to the full MoCA gold standard? To validate MoCA-SL, we examine its alignment with full MoCA scores across all participants with both assessments available. 
Figure~\ref{fig:score} plots MoCA-SL against full MoCA scores. 
Despite variance at low-score ranges due to sparse samples, the two measures exhibit a strong linear relationship (Pearson $r = 0.829$, $p < 0.001$). 
This correlation confirms that MoCA-SL preserves the discriminative validity of the full assessment.

\noindent \textbf{Verifier Max Retries.}
The Verification Loop iteratively corrects Examiner hallucinations, but excessive retries may introduce latency without performance gains. 
We analyze how the maximum retry $N_{max}$ limit affects examination accuracy. 
We vary $N_{max}$ from 0 to 3 and measure SMR on high-inference tasks where verification is most impactful. 
Figure~\ref{fig:score} (b) shows SMR results.
Performance improves sharply from $N_{max}$ = 0 to $N_{max}$ = 1, with diminishing returns beyond $N_{max}$ = 2. 
Based on this analysis, we set $N_{max}$ = 3 to balance accuracy and efficiency.

\begin{figure}[h]
\begin{center}
\includegraphics[width=0.85\linewidth,scale=1.0]{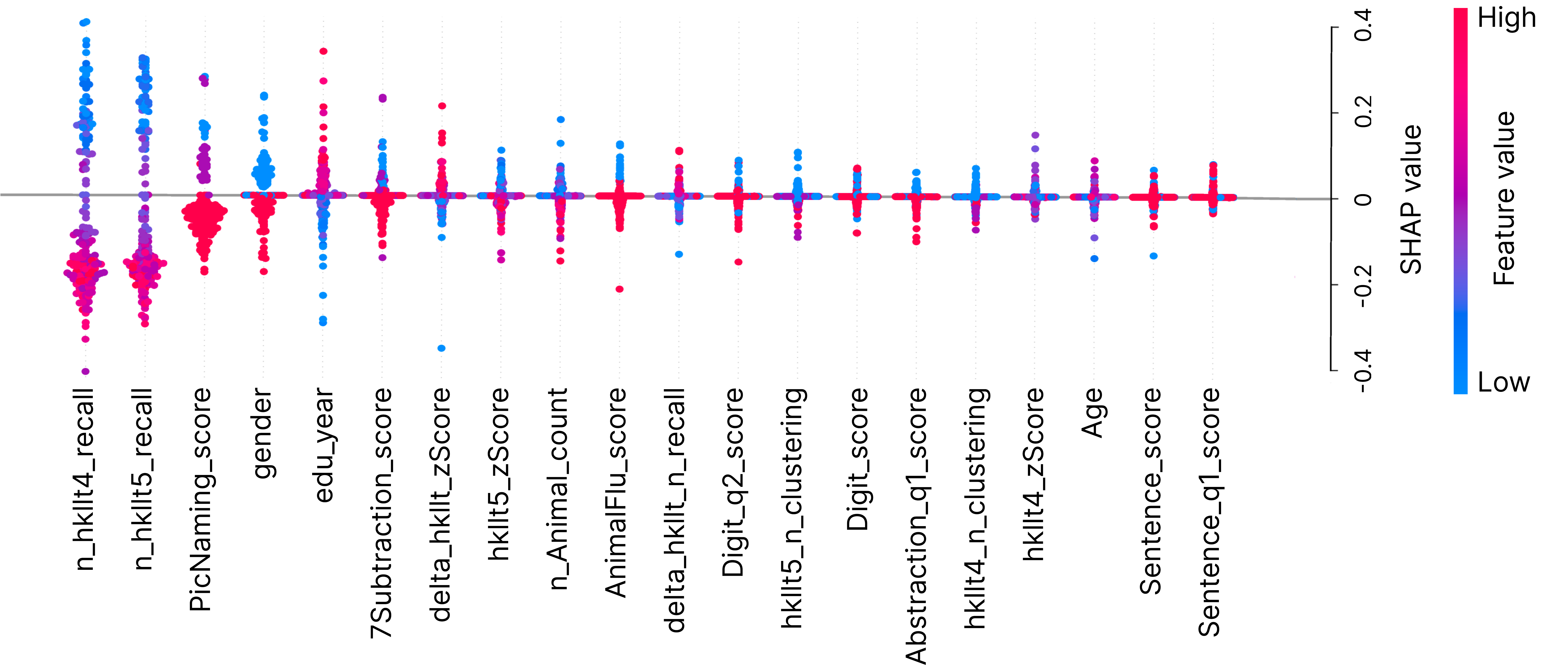}
\end{center}
\vspace{-15pt}
\caption{
Feature importance analysis using SHAP values for the supervised SVM classifier. 
Top-20 features are depicted.
}
\vspace{-5pt}
\label{fig:shap}
\end{figure}

\noindent \textbf{Feature Importance.}
We analyze SHAP values on the SVM classifier to interpret feature contributions. Figure~\ref{fig:shap} shows that memory-related HKLLT features (e.g., n\_hkllt4\_recall) are the most influential, consistent with clinical consensus that episodic memory is an effective early marker of AD risk.
MoCA-SL task scores (e.g., PicNaming, 7Subtraction) follow, aligning with clinical expectations of language and attention as secondary indicators. 
Notably, edu\_year shows a negative relationship—higher education is associated with reduced predicted AD risk, consistent with the cognitive reserve effect documented in clinical literature and the education-adjusted normative cutoffs used in clinical practice (see Appendix~\ref{appendix:norm}). 
Among low-impact features, we hypothesize that the Sentence task contributes minimally likely due to ceiling effects, and that HKLLT semantic clustering metrics may be masked by the dominant n\_recall signal.
\section{Conclusion}
\vspace{-10pt}
We presented an agentic cognitive assessment framework that bridges the gap between clinical protocol logic and automated Alzheimer's Disease screening. By decomposing standardized assessments into atomic cognitive tasks and delegating all quantification to deterministic functions, the framework restores construct validity to automated screening without sacrificing predictive performance.
Experiments on eight cognitive tasks and 402 participants confirm two key findings:
(a) the zero-shot setting already surpasses PLM-based baselines trained on task transcripts, suggesting that clinical protocol logic itself carries sufficient discriminative signal without requiring data-driven pattern extraction; (b) SHAP analysis confirms that the most influential features—notably episodic memory recall—align with established clinical markers of early AD, indicating that the extracted scoring primitives capture genuine cognitive constructs rather than statistical shortcuts.
Together, these results demonstrate that construct validity and predictive performance are complementary rather than competing objectives, offering a principled foundation for automated cognitive assessment systems that explain rather than merely predict.



\section*{Limitations}

\textit{1. Dependence on Predefined Scoring Rules}

\noindent The framework’s operation aligns with standardized clinical protocols, and its validity is therefore contingent on the implementation of established scoring rules. While this ensures fidelity to clinical constructs, the assessment is inherently confined to the cognitive domains predefined by the tasks.

\noindent\textit{2. Reliance on LLM Semantic Capability}

\noindent While deterministic functions ensure objective quantification, the agents' semantic parsing performance remains limited by the LLM's comprehension capabilities.

\noindent\textit{3. Restricted Dataset Accessibility}

\noindent Due to Institutional Review Board constraints and ethical guidelines, the clinical corpus used in this work cannot be shared publicly, limiting direct reproducibility.
However, the framework's architecture is not language-specific. Adapting it to other languages requires only re-authoring task-specific prompts and normative tables, while the overall pipeline structure remains transferable.

\section*{Ethical Considerations}

We used LLMs (e.g., Claude and Gemini) to assist with language editing and clarification during the preparation of this paper. The use of LLMs was limited to improving readability and expression, and all technical content, ideas, and research contributions are solely those of the authors. We confirm compliance with ACL ethical guidelines.

\section*{Reproducibility Statement}
All prompts and scoring criteria are provided in Appendices B–E. The assessment protocol and feature extraction procedures are detailed in Sections 3–4 and Appendix A. The clinical dataset cannot be publicly released due to IRB constraints on protected health information, but we provide sufficient methodological detail to enable replication with independently collected data.

\section*{Acknowledgments}

This work is supported by the HKSARG Research Grants Council’s Theme-based Research Grant Scheme (Project No. T45-407/19N), the CUHK Stanley Ho Big Data Decision Research Centre, and the Centre for Perceptual and Interactive Intelligence (CPII) Ltd, a CUHK-led InnoCentre under the InnoHK initiative of the Innovation and Technology Commission of the Hong Kong Special Administrative Region Government.

\bibliography{custom}

@inproceedings{park2025reasoning,
  title={Reasoning-Based Approach with Chain-of-Thought for Alzheimer's Detection Using Speech and Large Language Models},
  author    = {Chanwoo Park and Anna Seo Gyeong Choi and Sunghye Cho and Chanwoo Kim},
  year      = {2025},
  booktitle = {{Interspeech 2025}},
  pages     = {2185--2189},
  doi       = {10.21437/Interspeech.2025-1226},
  issn      = {2958-1796},
}

@article{li2025detecting,
  title={Detecting neurocognitive disorders through analyses of topic evolution and cross-modal consistency in visual-stimulated narratives},
  author={Li, Jinchao and Wang, Yuejiao and Li, Junan and Kang, Jiawen and Zheng, Bo and Wong, Ka Ho and Mak, Brian and Fung, Helene H and Woo, Jean and Mak, Man-Wai and others},
  journal={IEEE Journal of Selected Topics in Signal Processing},
  year={2025},
  publisher={IEEE},
  doi={10.1109/JSTSP.2025.3622049},
}

@inproceedings{li2024devising,
  title={Devising a Set of Compact and Explainable Spoken Language Feature for Screening Alzheimer's Disease},
  author={Li, Junan and Li, Yunxiang and Wang, Yuren and Wu, Xixin and Meng, Helen},
  booktitle={2024 IEEE 14th International Symposium on Chinese Spoken Language Processing (ISCSLP)},
  pages={471--475},
  year={2024},
  organization={IEEE},
  doi={10.1109/ISCSLP63861.2024.10800414}
}

@misc{bazgir2025agent,
      title={AgenticAD: A Specialized Multiagent System Framework for Holistic Alzheimer Disease Management}, 
      author={Adib Bazgir and Amir Habibdoust and Xing Song and Yuwen Zhang},
      year={2025},
      eprint={2510.08578},
      archivePrefix={arXiv},
      primaryClass={cs.MA},
      url={https://arxiv.org/abs/2510.08578}, 
}

@inproceedings{hou2025adagent,
  title = {{{ADAgent}}: {{LLM}} Agent for Alzheimer's Disease Analysis with Collaborative Coordinator},
  booktitle = {{{AI}} for Clinical Applications},
  author = {Hou, Wenlong and Yang, Guangqian and Du, Ye and Lau, Yeung and Liu, Lihao and He, Junjun and Long, Ling and Wang, Shujun},
  editor = {Qiu, Jianing and Wu, Jinlin and Langlotz, Curtis and Huang, Baoru and Lei, Zhen and Wu, Honghan and Liu, Hongbin and Xie, Weidi},
  year = 2026,
  pages = {23--32},
  url = {https://doi.org/10.1007/978-3-032-06004-4_3},
  publisher = {Springer Nature Switzerland},
  isbn = {978-3-032-06004-4}
}

@article{li2025care,
  title={CARE-AD: a multi-agent large language model framework for Alzheimer’s disease prediction using longitudinal clinical notes},
  author={Li, Rumeng and Wang, Xun and Berlowitz, Dan and Mez, Jesse and Lin, Honghuang and Yu, Hong},
  journal={npj Digital Medicine},
  volume={8},
  number={1},
  pages={541},
  year={2025},
  publisher={Nature Publishing Group UK London},
  doi={https://doi.org/10.1038/s41746-025-01940-4}
}

@inproceedings{ding2022talktive,
  title={TalkTive: A conversational agent using backchannels to engage older adults in neurocognitive disorders screening},
  author={Ding, Zijian and Kang, Jiawen and Ho, Tinky Oi Ting and Wong, Ka Ho and Fung, Helene H and Meng, Helen and Ma, Xiaojuan},
  booktitle={Proceedings of the 2022 CHI conference on human factors in computing systems},
  pages={1--19},
  year={2022},
  doi={10.1145/3491102.3502005}
}

@misc{an2025conversational,
      title={Conversational Agents for Older Adults' Health: A Systematic Literature Review}, 
      author={Jiaxin An and Siqi Yi and Yao Lyu and Houjiang Liu and Yan Zhang},
      year={2025},
      eprint={2503.23153},
      archivePrefix={arXiv},
      primaryClass={cs.HC},
      url={https://arxiv.org/abs/2503.23153}, 
}

@misc{breithaupt2025designing,
      title={Designing and Evaluating a Conversational Agent for Early Detection of Alzheimer's Disease and Related Dementias}, 
      author={Andrew G. Breithaupt and Nayoung Choi and James D. Finch and Jeanne M. Powell and Arin L. Nelson and Oz A. Alon and Howard J. Rosen and Jinho D. Choi},
      year={2025},
      eprint={2509.11478},
      archivePrefix={arXiv},
      primaryClass={cs.HC},
      url={https://arxiv.org/abs/2509.11478}, 
}

@inproceedings{yang2024sweagent,
  title={{SWE}-agent: Agent-Computer Interfaces Enable Automated Software Engineering},
  author={John Yang and Carlos E Jimenez and Alexander Wettig and Kilian Lieret and Shunyu Yao and Karthik R Narasimhan and Ofir Press},
  booktitle={The Thirty-eighth Annual Conference on Neural Information Processing Systems},
  year={2024},
  url={https://arxiv.org/abs/2405.15793}
}

@article{m2024augmenting,
  title={Augmenting large language models with chemistry tools},
  author={M. Bran, Andres and Cox, Sam and Schilter, Oliver and Baldassari, Carlo and White, Andrew D and Schwaller, Philippe},
  journal={Nature Machine Intelligence},
  volume={6},
  number={5},
  pages={525--535},
  year={2024},
  publisher={Nature Publishing Group UK London},
  doi={https://doi.org/10.1038/s42256-024-00832-8},
}

@article{wang2023voyager,
  title={Voyager: An open-ended embodied agent with large language models},
  author={Wang, Guanzhi and Xie, Yuqi and Jiang, Yunfan and Mandlekar, Ajay and Xiao, Chaowei and Zhu, Yuke and Fan, Linxi and Anandkumar, Anima},
  journal={Transactions on Machine Learning Research},
  issn={2835-8856},
  year={2024},
  url={https://openreview.net/forum?id=ehfRiF0R3a},
}

@article{hobson2015montreal,
  title={The montreal cognitive assessment (MoCA)},
  author={Hobson, John},
  journal={Occupational Medicine},
  volume={65},
  number={9},
  pages={764--765},
  year={2015},
  publisher={Oxford University Press UK},
  doi={10.1093/occmed/kqv078},
}

@article{yeung2014validation,
  title={A validation study of the Hong Kong version of Montreal Cognitive Assessment (HK-MoCA) in Chinese older adults in Hong Kong},
  author={Yeung, PY and Wong, LL and Chan, CC and Leung, Jess LM and Yung, CY and others},
  journal={Hong Kong Medical Journal},
  year={2014},
  doi={10.12809/hkmj144219}
}

@article{delis2000california,
  title={California verbal learning test--Second Edition},
  author={Delis, Dean C and Kramer, Joel H and Kaplan, Edith and Ober, Beth A},
  journal={Assessment},
  year={2000},
  doi={10.1037/t15072-000}
}

@article{au2003verbal,
  title={Verbal learning in Alzheimer's dementia},
  author={Au, Alma and Chan, Agnes S and Chiu, Helen},
  journal={Journal of the International Neuropsychological Society},
  volume={9},
  number={3},
  pages={363--375},
  year={2003},
  doi={10.1017/S1355617703930025}
}

@article{giles1996performance,
  title={Performance on the Boston Cookie Theft picture description task in patients with early dementia of the Alzheimer's type: missing information},
  author={Giles, Elaine and Patterson, Karalyn and Hodges, John R},
  journal={Aphasiology},
  volume={10},
  number={4},
  pages={395--408},
  year={1996},
  publisher = {Routledge},
  doi={10.1080/02687039608248419}
}

@article{fraser2016linguistic,
  title={Linguistic features identify Alzheimer’s disease in narrative speech},
  author={Fraser, Kathleen C and Meltzer, Jed A and Rudzicz, Frank},
  journal={Journal of Alzheimer's Disease},
  year={2016},
  publisher={IOS Press},
  volume = {49},
  number = {2},
  pages = {407--422},
  doi={10.3233/JAD-150520}
}

@inproceedings{weiner2019speech,
  title={Speech reveals future risk of developing dementia: Predictive dementia screening from biographic interviews},
  author={Weiner, Jochen and Frankenberg, Claudia and Schr{\"o}der, Johannes and Schultz, Tanja},
  booktitle={2019 IEEE Automatic Speech Recognition and Understanding Workshop (ASRU)}, 
  year={2019},
  volume={},
  number={},
  pages={674-681},
  doi={10.1109/ASRU46091.2019.9003908}}

@inproceedings{balagopalan2020bert,
  title     = {To BERT or not to BERT: Comparing Speech and Language-Based Approaches for Alzheimer’s Disease Detection},
  author    = {Aparna Balagopalan and Benjamin Eyre and Frank Rudzicz and Jekaterina Novikova},
  year      = {2020},
  booktitle = {Interspeech 2020},
  pages     = {2167--2171},
  doi       = {10.21437/Interspeech.2020-2557},
  issn      = {2958-1796},
}

@inproceedings{yuan2020disfluencies,
  title={Disfluencies and Fine-Tuning Pre-Trained Language Models for Detection of Alzheimer's Disease},
  author    = {Jiahong Yuan and Yuchen Bian and Xingyu Cai and Jiaji Huang and Zheng Ye and Kenneth Church},
  year      = {2020},
  booktitle = {Interspeech 2020},
  pages     = {2162--2166},
  doi       = {10.21437/Interspeech.2020-2516},
  issn      = {2958-1796},
}

@inproceedings{adress,
  title={{Alzheimer’s Dementia Recognition Through Spontaneous Speech: The ADReSS Challenge}},
  author    = {Saturnino Luz and Fasih Haider and Sofia de la Fuente and Davida Fromm and Brian MacWhinney},
  year      = {2020},
  booktitle = {Interspeech 2020},
  pages     = {2172--2176},
  doi       = {10.21437/Interspeech.2020-2571},
  issn      = {2958-1796},
}

@inproceedings{adresso,
  title={Detecting cognitive decline using speech only: The ADReSSo Challenge},
  author    = {Saturnino Luz and Fasih Haider and Sofia de la Fuente and Davida Fromm and Brian MacWhinney},
  year      = {2021},
  booktitle = {Interspeech 2021},
  pages     = {3780--3784},
  doi       = {10.21437/Interspeech.2021-1220},
  issn      = {2958-1796},
}

@inproceedings{liu2024clever,
  title={Clever hans effect found in automatic detection of alzheimer's disease through speech},
  author    = {Yin-Long Liu and Rui Feng and Jia-Hong Yuan and Zhen-Hua Ling},
  year      = {2024},
  booktitle = {{Interspeech 2024}},
  pages     = {2435--2439},
  doi       = {10.21437/Interspeech.2024-1018},
  issn      = {2958-1796},
}

@article{sahidullah2025shortcut,
  title={Shortcut Learning in Binary Classifier Black Boxes: Applications to Voice Anti-Spoofing and Biometrics},
  author={Sahidullah, Md and Shim, Hye-jin and Hautamäki, Rosa Gonzalez and Kinnunen, Tomi H.},
  journal={IEEE Journal of Selected Topics in Signal Processing},
  year={2025},
  publisher={IEEE},
  doi={10.1109/JSTSP.2025.3569430},
}

@inproceedings{kang2024within,
  title     = {{On the Within-class Variation Issue in Alzheimer's Disease Detection}},
  author    = {Jiawen Kang and Dongrui Han and Lingwei Meng and Jingyan Zhou and Jinchao Li and Xixin Wu and Helen Meng},
  year      = {2025},
  booktitle = {{Interspeech 2025}},
  pages     = {5668--5672},
  doi       = {10.21437/Interspeech.2025-2751},
  issn      = {2958-1796},
}

@inproceedings{kang2024not,
  title={Not All Errors Are Equal: Investigation of Speech Recognition Errors in Alzheimer's Disease Detection},
  author={Kang, Jiawen and Li, Junan and Li, Jinchao and Wu, Xixin and Meng, Helen},
  booktitle={2024 IEEE 14th International Symposium on Chinese Spoken Language Processing (ISCSLP)},
  pages={254--258},
  year={2024},
  organization={IEEE},
  doi={10.1109/ISCSLP63861.2024.10800483}}

@article{li2024useful,
  title={Useful blunders: Can automated speech recognition errors improve downstream dementia classification?},
  author={Li, Changye and Xu, Weizhe and Cohen, Trevor and Pakhomov, Serguei},
  journal={Journal of biomedical informatics},
  volume={150},
  pages={104598},
  year={2024},
  publisher={Elsevier},
  issn = {1532-0464},
  doi = {10.1016/j.jbi.2024.104598},
}

@article{strauss2009construct,
  title={Construct validity: Advances in theory and methodology},
  author={Strauss, Milton E and Smith, Gregory T},
  journal={Annual review of clinical psychology},
  volume={5},
  pages={1--25},
  year={2009},
  publisher={Annual Reviews},
  doi={10.1146/annurev.clinpsy.032408.153639}
}

@article{van2024undesirable,
  title={Undesirable biases in NLP: Addressing challenges of measurement},
  author={Van der Wal, Oskar and Bachmann, Dominik and Leidinger, Alina and Van Maanen, Leendert and Zuidema, Willem and Schulz, Katrin},
  journal={Journal of Artificial Intelligence Research},
  volume={79},
  pages={1--40},
  year={2024},
  doi = {10.1613/jair.1.15195},
}

@article{harvey2012clinical,
  title={Clinical applications of neuropsychological assessment},
  author={Harvey, Philip D},
  journal={Dialogues in clinical neuroscience},
  volume={14},
  number={1},
  pages={91--99},
  year={2012},
  publisher={Taylor \& Francis},
  doi={10.31887/DCNS.2012.14.1/pharvey}
}

@article{galvin2005ad8,
  title={The AD8: a brief informant interview to detect dementia},
  author={Galvin, JE and Roe, CM and Powlishta, KK and Coats, MA and Muich, SJ and Grant, E and Miller, JP and Storandt, M and Morris, JC},
  journal={Neurology},
  volume={65},
  number={4},
  pages={559--564},
  year={2005},
  publisher={Lippincott Williams \& Wilkins},
  doi={10.1212/01.wnl.0000172958.95282.2a}
}

@book{goodglass2001bdae,
  title={BDAE: The Boston diagnostic aphasia examination},
  author={Goodglass, Harold and Kaplan, Edith and Weintraub, Sandra},
  year={2001},
  publisher={Lippincott Williams \& Wilkins Philadelphia, PA},
  url={https://utdallas.primo.exlibrisgroup.com/discovery/fulldisplay/alma9927692213901421/01UT_DALLAS:UTDALMA}
}

@article{cockrell2002mini,
  title={Mini-mental state examination},
  author={Cockrell, Joseph R and Folstein, Marshal F},
  journal={Principles and practice of geriatric psychiatry},
  pages={140--141},
  year={2002},
  publisher={Wiley Online Library},
  doi={10.1002/0470846410.ch27(ii)}
}

@inproceedings{pipeline,
  title={Integrated and enhanced pipeline system to support spoken language analytics for screening neurocognitive disorders},
  author    = {Helen Meng and Brian Mak and Man-Wai Mak and Helene Fung and Xianmin Gong and Timothy Kwok and Xunying Liu and Vincent Mok and Patrick Wong and Jean Woo and Xixin Wu and Ka Ho Wong and Shensheng Xu and Naijun Zheng and Ranzo Huang and Jiawen Kang and Xiaoquan Ke and Junan Li and Jinchao Li and Yi Wang},
  year      = {2023},
  booktitle = {Interspeech 2023},
  pages     = {1713--1717},
  doi       = {10.21437/Interspeech.2023-2249},
  issn      = {2958-1796},
}

@article{haulcy2021classifying,
  title={Classifying Alzheimer's disease using audio and text-based representations of speech},
  author={Haulcy, R'mani and Glass, James},
  journal={Frontiers in Psychology},
  volume={11},
  pages={624137},
  year={2021},
  publisher={Frontiers Media SA},
  doi={10.3389/fpsyg.2020.624137}
}

@inproceedings{jinchao23,
  title={Leveraging pretrained representations with task-related keywords for alzheimer’s disease detection},
  author={Li, Jinchao and Song, Kaitao and Li, Junan and Zheng, Bo and Li, Dongsheng and Wu, Xixin and Liu, Xunying and Meng, Helen},
  booktitle={ICASSP 2023-2023 IEEE International Conference on Acoustics, Speech and Signal Processing (ICASSP)},
  pages={1--5},
  year={2023},
  organization={IEEE},
  doi={10.1109/ICASSP49357.2023.10096205}
}

@inproceedings{koo2020exploiting,
  title     = {Exploiting Multi-Modal Features from Pre-Trained Networks for Alzheimer’s Dementia Recognition},
  author    = {Junghyun Koo and Jie Hwan Lee and Jaewoo Pyo and Yujin Jo and Kyogu Lee},
  year      = {2020},
  booktitle = {Interspeech 2020},
  pages     = {2217--2221},
  doi       = {10.21437/Interspeech.2020-3153},
  issn      = {2958-1796},
}

@article{syed2021automated,
  title={Automated recognition of Alzheimer’s dementia using bag-of-deep-features and model ensembling},
  author={Syed, Zafi Sherhan and Syed, Muhammad Shehram Shah and Lech, Margaret and Pirogova, Elena},
  journal={IEEE Access},
  year={2021},
  publisher={IEEE},
  doi={10.1109/ACCESS.2021.3090321}
}

@inproceedings{wavbert,
  title={Wavbert: Exploiting semantic and non-semantic speech using wav2vec and bert for dementia detection},
  author={Zhu, Youxiang and Obyat, Abdelrahman and Liang, Xiaohui and Batsis, John A and Roth, Robert M},
  year      = {2021},
  booktitle = {Interspeech 2021},
  pages     = {3790--3794},
  doi       = {10.21437/Interspeech.2021-332},
  issn      = {2958-1796},
}

@inproceedings{yi22,
  title={Exploring linguistic feature and model combination for speech recognition based automatic ad detection},
  author    = {Wang, Yi and Wang, Tianzi and Ye, Zi and Meng, Lingwei and Hu, Shoukang and Wu, Xixin and Liu, Xunying and Meng, Helen},
  year      = {2022},
  booktitle = {Interspeech 2022},
  pages     = {3328--3332},
  doi       = {10.21437/Interspeech.2022-723},
  issn      = {2958-1796},
}

@article{wong2015montreal,
  title={Montreal cognitive assessment: one cutoff never fits all},
  author={Wong, Adrian and Law, Lorraine SN and Liu, Wenyan and Wang, Zhaolu and Lo, Eugene SK and Lau, Alexander and Wong, Lawrence KS and Mok, Vincent CT},
  journal={Stroke},
  volume={46},
  number={12},
  pages={3547--3550},
  year={2015},
  publisher={Lippincott Williams \& Wilkins Hagerstown, MD},
  doi={10.1161/STROKEAHA.115.011226}
}
\bibliographystyle{iclr2026_conference}
\newpage
\appendix

\section{Detailed Description of Cognitive Assessments}
\label{appendix:protocol}
\subsection{Montreal Cognitive Assessment (MoCA)}
Table \ref{tab:moca} presents a summary of the Montreal Cognitive Assessment (MoCA), a widely used screening instrument designed to assess mild cognitive dysfunction by probing multiple cognitive domains, including attention, executive functions, memory, language, and conceptual thinking. 
In this work, we utilized a specific subset termed MoCA-SL (Spoken Language), which is highlighted with an asterisk (*) in the table. 
These tasks—including Picture Naming, Digit Span, Serial 7 Subtraction, Sentence Repetition, Verbal Fluency, and Abstraction—are based entirely on verbal interaction. 
Unlike visual-motor tasks (e.g., trail making), these spoken tasks could be readily recorded and processed by an automated pipeline.

\begin{table*}[h]
\centering
\renewcommand{\arraystretch}{1.5} %
\caption{Detailed description and scoring criteria for the Montreal Cognitive Assessment (MoCA, Hong Kong version).
Tasks marked with * belong to the spoken-language subset (MoCA-SL) used in this study
}
\label{tab:moca}
\scalebox{0.7}{
\begin{tabular}{lp{5.7cm}p{3.2cm}p{4.1cm}r}
\toprule
\textbf{Task} & \textbf{Description} & \makecell[c]{\textbf{Cognitive Domain} \\ \textbf{(DSM-5)}} & \textbf{Scoring Primitives} & \textbf{Score} \\

\midrule
Alternating Trail Making & 
Draw a line to connect numbers and characters in alternating order & 
Exec. Function & 
Successful connection pattern & 1 \\

Copy Cube & 
Copy a 3-dimensional cube structure & 
Perceptual Motor & 
Correct 3D structure & 1 \\

Clock Drawing & 
Draw a clock face, numbers, and set time to 11:10 & 
Perceptual Motor & 
Contour (1), Numbers (1), Hands (1) & 3 \\

*Picture Naming & 
Name three depicted animals (e.g., Lion, Rhino, Camel) & 
Language & 
Per-item correctness ($\times$3) & 3 \\

*Digit Span & 
Repeat digits forward and backward & 
Complex Attention & 
Forward (1), Backward (1) & 2 \\

Vigilance (Tap) & 
Tap hand at each target digit (e.g., '1') in a sequence & 
Complex Attention & 
$\le$ 1 error allowed & 1 \\

*Serial 7 Subtraction & 
Subtract 7 from 100 consecutively & 
Complex Attention & 
4-5 correct (3), 2-3 correct (2), 1 correct (1) & 3 \\

*Sentence Repetition & 
Repeat two sentences verbatim & 
Language & 
Per-sentence correctness ($\times$2) & 2 \\

*Animal Fluency & 
Generate animal words starting with a specific character/category in 1 min & 
Language, Execution Function & 
$\ge$ threshold (e.g., 11 words) & 1 \\

*Abstraction & 
Identify similarity between word pairs & 
Executive Function & 
Per-pair correctness ($\times$2) & 2 \\

Delayed Recall & 
Recall 5 words learned earlier (after $\sim$5 min) & 
Learning and Memory & 
Per-word correctness ($\times$5) & 5 \\

Orientation & 
Identify current date, month, year, day, place, and city & 
Perceptual Motor, Orientation & 
Per-item correctness ($\times$6) & 6 \\

& & & \multicolumn{2}{r}{\textit{MoCA Total:} 30}\\

\bottomrule
\end{tabular}}
\end{table*}

\subsection{Hong Kong List Learning}
\begin{table*}[t]
\centering
\renewcommand{\arraystretch}{1.5} %
\caption{Detailed description and scoring criteria for the Hong Kong List Learning Test (HKLLT)
}
\label{tab:hkllt_tasks}
\scalebox{0.75}{
\begin{tabular}{p{2.3cm}p{5cm}p{3cm}p{5.5cm}c}
\toprule
\textbf{Trial / Task} & \textbf{Description} & \makecell[c]{\textbf{Cognitive Domain} \\ \textbf{(DSM-5)}} & \textbf{Scoring Primitives} & \textbf{Score} \\
\midrule
\multicolumn{5}{l}{\textit{Acquisition \& Learning (Form 1: Random Condition)}} \\
\midrule
Trial 1 & 
Listen to a list of 16 words (4 categories) presented orally, followed by immediate free recall. & 
Immediate Memory & 
\# correctly recalled words & 16 \\

Trials 2 \& 3 & 
Listen to the same list repeated (same order), followed by free recall each time. & 
Learning Rate & 
\# correctly recalled words (per trial) & 16 \\

\midrule
\multicolumn{5}{l}{\textit{Delayed Recall (Retention)}} \\
\midrule
Trial 4 \newline(10-min Delay) & 
Unannounced free recall of the 16 words after a 10-minute non-verbal interval. & 
Short-term Retention & 
\# correctly recalled words & 16 \\

Trial 5 \newline (30-min Delay) & 
Unannounced free recall of the 16 words after a 30-minute interval. & 
Long-term Retention & 
\# correctly recalled words & 16 \\

\midrule
\multicolumn{5}{l}{\textit{Recognition}} \\
\midrule
Recognition & 
Identify the 16 target words mixed with 16 distractors (foils) presented orally (Yes/No). & 
Recognition Memory & 
Hits (True Positives) \& False Alarms. 
\newline \textit{Discrim. Score} = $\frac{\text{Hits} - \text{False Alarms}}{16} \times 100\%$ & 16 \\

\bottomrule
\end{tabular}}
\end{table*}

Table \ref{tab:hkllt_tasks} presents a summary of the Hong Kong List Learning Test (HKLLT), a standardized neuropsychological assessment that mainly probes verbal learning and episodic memory through the recall of a 16-word list. In this work, we specifically focused on Trial 4 (10-minute delayed recall) and Trial 5 (30-minute delayed recall). These delayed recall measures are regarded as particularly effective for early diagnosis, as clinical research demonstrates that the rate of rapid forgetting over these intervals is the most sensitive discriminator between patients with mild Alzheimer’s disease and normal elderly controls.






\begin{table*}[h]
\centering
\renewcommand{\arraystretch}{1.2} %
\caption{Handcraft linguistic features used in this work.
``\%'': ``ratio of'', ``\#'': ``number of'', ``\textnormal{dur}'': ``duration''.}
\label{tab:biomarkers}
\setlength\tabcolsep{3mm}
\resizebox{0.55\linewidth}{!}{
\begin{tabular}{l l l}
\toprule
\textbf{ID} & \textbf{Feature Name} & \textbf{Description}                            \\
\midrule
L1          & \# words & The number of words                     \\
L2          & \% Stop words & \# stop words / \# words                     \\
L3          & \% Filled pauses & \# filler words / \# words                \\
L4          & \% Lexical filler & \# lexical filler  / \# words            \\
L5          & \% Backchannel & \# backchannel words  / \# words            \\
L6          & \% Repetition & \# repeated words / \# words                 \\
L7         & \% Adj & \# adjectives / \# words                            \\
L8         & \% Adv & \# adverbs / \# words                               \\
L9          & \% Noun & \# nouns / \# words                               \\
L10          & \% Pronoun & \# pronouns / \# words                          \\
L11          & \% Verb & \# verb / \# words                                 \\
L12         & \% Func & \# functional words / \# words                               \\
L13         & \% CTTR & \# types / rooted(2 * \# tokens)     \\
\bottomrule
\end{tabular}
}
\vspace{-0.5em}
\end{table*}

\section{Implementation Details}
\label{appendix:implementation}
\paragraph{Model \& Deployment.}
We use Qwen3-8B\footnote{
https://huggingface.co/Qwen/Qwen3-8B} as the backbone for all agents, deployed locally via vLLM with half precision and 0.85 GPU memory utilization. 
All experiments are conducted on 4 NVIDIA RTX 8000 (48GB).

\paragraph{Agent Configuration.}
Key inference parameters include: max sequence length 4096, temperature 0.3 (Examiner) / 0.1 (Verifier), and top\_p 0.9.

\paragraph{Baselines.}
Handcrafted linguistic features are detailed in Table~\ref{tab:biomarkers}. For PLM-based methods, we use BERT\footnote{https://huggingface.co/google-bert/bert-base-chinese} and RoBERTa\footnote{https://huggingface.co/hfl/chinese-roberta-wwm-ext} with transcripts translated to Simplified Chinese. 
We do not finetune these models, as preliminary experiments showed no consistent improvement. 
Task responses are concatenated with "[$<$TASK$>$-$<$NAME$>$]" delimiters as input. 
For LLM-COT, the prompt template is provided in Appendix~\ref{appendix:prompts}.

\paragraph{Classifiers.}
SVM uses RBF kernel with C=1.0 and gamma=scale. 
MLP consists of two hidden layers (32, 16) with ReLU activation, trained for 20 epochs with dropout 0.2, batch size 16, and initial learning rate 0.001.

\paragraph{Scoring Primitives.}
The complete list of scoring primitives used as classifier input is provided in Table~\ref{tab:primitives} .

\begin{table*}[tbh]
\centering
\renewcommand{\arraystretch}{1.2} %
\caption{Cognitive scoring primitives used in this work.
``\#'': ``number of'', ``z-score'': ``standardized score relative to norms''.}
\label{tab:primitives}

\setlength\tabcolsep{3mm}
\resizebox{0.75\linewidth}{!}{
\begin{tabular}{l l l}
\toprule
\textbf{ID} & \textbf{Feature Name} & \textbf{Description} \\
\toprule
\multicolumn{3}{l}{\textit{HKLLT (Hong Kong List Learning Test)}} \\
\midrule
C1  & hkllt4\_zScore & Trial 4 z-score \\
C2  & hkllt5\_zScore & Trial 5 z-score \\
C3  & n\_hkllt4\_recall & Trial 4 recall count \\
C4  & n\_hkllt5\_recall & Trial 5 recall count \\
C5  & hkllt4\_n\_clustering & Trial 4 semantic clustering count \\
C6  & hkllt5\_n\_clustering & Trial 5 semantic clustering count \\
C7  & delta\_hkllt\_zScore & Z-score difference between Trial 4 and Trial 5 \\
C8  & delta\_hkllt\_n\_recall & Recall count difference between Trial 4 and Trial 5 \\
\toprule
\multicolumn{3}{l}{\textit{MoCA-SL (Montreal Cognitive Assessment)}} \\
\midrule
M1  & n\_Animal\_count & Animal fluency count \\
M2  & AnimalFlu\_score & Animal fluency score \\
M3  & 7Subtraction\_score & Serial 7s subtraction score \\
M4  & Abstraction\_q1\_score & Abstraction item 1 score \\
M5  & Abstraction\_q2\_score & Abstraction item 2 score \\
M6  & Digit\_fwd\_score & Digit span forward score \\
M7  & Digit\_bwd\_score & Digit span backward score \\
M8  & Digit\_score & Digit span total score \\
M9  & PicNaming\_score & Picture naming score \\
M10 & Sentence\_q1\_score & Sentence repetition item 1 score \\
M11 & Sentence\_q2\_score & Sentence repetition item 2 score \\
M12 & Sentence\_score & Sentence repetition total score \\
\bottomrule
\end{tabular}
}
\end{table*}
\section{Population Norms for MoCA-SL Scores}
\label{appendix:norm}

Since MoCA-SL comprises only a subset of the full MoCA assessment, existing normative tables cannot be directly applied. 
To enable norm-referenced scoring and percentile calculation, we explored three approaches to estimate population norms for MoCA-SL.

\paragraph{Approach 1: Direct Empirical Estimation.}
The most straightforward approach involves computing MoCA-SL norms directly from healthy control participants in our corpus. 
We stratified 536 cognitively normal participants by age and education, then calculated mean and standard deviation for each stratum. 
While this method is intuitive, it carries the risk that our sample may not adequately represent the broader Hong Kong elderly population, particularly in undersampled demographic cells.

\paragraph{Approach 2: Proportional Rescaling.}
This method leverages the established MoCA normative table~\citep{wong2015montreal} by rescaling all values proportionally.
Given that MoCA-SL covers 13 out of 30 total points, we apply a linear scaling factor of 13/30 to all norm values in the published table:
$$Z_{\text{norm}} = Y_{\text{norm}} \times \frac{13}{30}$$
where $Y_{\text{norm}}$ denotes norm values from the full MoCA table and $Z_{\text{norm}}$ represents the estimated MoCA-SL norms. 
This approach assumes that the subset score distribution maintains proportional relationships with the full score, ensuring full grounding in the clinically validated gold standard.

\paragraph{Approach 3: Regression-Based Estimation. }
We adopt a data-driven approach by fitting a linear regression model on our corpus to characterize the relationship between full MoCA scores ($Y$) and MoCA-SL scores ($Z$):
$$Z = \alpha + \beta \cdot Y$$
After estimating coefficients $\alpha$ and $\beta$ from healthy control participants, we apply this mapping to the published MoCA normative values to derive corresponding MoCA-SL norms:
$$Z_{\text{norm}} = \alpha + \beta \cdot Y_{\text{norm}}$$
This method combines empirical data with established norms, theoretically providing accurate estimates anchored to the gold standard while accounting for any non-linear ceiling or floor effects in the subset.

\paragraph{Experiments}
We conduct experiments to compare AD screening performance using different approaches; the results are shown in Table~\ref{tab:norm_comparison}.
We finally adopt Approach 2 (Proportional Rescaling) as the primary method due to its simplicity and superior performance.
The obtained normative table is shown in Table~\ref{tab:norm}.

\begin{table}[h]
\renewcommand{\arraystretch}{1.2} %
\centering
\caption{Comparison of AD screening performance using different MoCA-SL normative estimation approaches.}
\label{tab:norm_comparison}
\scalebox{0.85}{
\begin{tabular}{ccccc}
\hline
\textbf{Approach} & \textbf{Accuracy} & \textbf{Precision} & \textbf{Recall} & \textbf{F1} \\
\hline
1 & 0.818 & 0.813 & 0.786 & 0.796 \\
2 & 0.824 & 0.814 & 0.802 & 0.802 \\
3 & 0.818 & 0.813 & 0.786 & 0.796 \\
\hline
\end{tabular}}
\end{table}

\begin{table}[htbp]
\renewcommand{\arraystretch}{1.2} %
\centering
\caption{Normative Table for MoCA-SL scores using proportional rescaling.}
\scalebox{0.85}{
\label{tab:norm}
\small
\begin{tabular}{ll ccccc cc}
\toprule
\textbf{Age} & \textbf{Education} & \textbf{N} & \textbf{Median} & \textbf{IQR} & \multicolumn{3}{c}{\textbf{Percentile}} \\
\cmidrule(lr){6-8}
(years) & (years) & & & & 16th & 7th & 2nd \\
\midrule
65--69 & 0--3   & 64 & 9.1  & 1.7 & 7.4  & 6.1  & 3.9 \\
       & 4--6   & 82 & 10.0 & 2.2 & 8.2  & 7.8  & 5.6 \\
       & 7--9   & 74 & 10.4 & 2.2 & 9.1  & 8.2  & 6.9 \\
       & 10--12 & 82 & 10.8 & 1.7 & 9.5  & 8.7  & 7.4 \\
       & $>$12  & 67 & 11.7 & 1.3 & 10.8 & 10.0 & 9.1 \\
\addlinespace
70--79 & 0--3   & 76 & 8.2  & 2.2 & 6.5  & 6.1  & 4.8 \\
       & 4--6   & 82 & 9.5  & 1.7 & 7.8  & 6.5  & 4.3 \\
       & 7--9   & 66 & 10.0 & 1.7 & 8.7  & 7.8  & 6.5 \\
       & 10--12 & 76 & 10.4 & 1.7 & 9.5  & 8.2  & 7.8 \\
       & $>$12  & 67 & 10.8 & 2.2 & 9.5  & 8.7  & 6.9 \\
\addlinespace
$\ge$80 & 0--6   & 37 & 7.8  & 2.6 & 5.6  & 5.6  & 4.3 \\
       & $>$6   & 21 & 8.7  & 2.2 & 7.4  & 6.5  & 5.6 \\
\bottomrule
\end{tabular}}
\end{table}
\section{Task Examination Evaluation Details}
\label{appendix:task_eval_details}

\begin{table}[t]
\centering
\caption{
Full metric breakdown for MoCA-SL and HKLLT tasks.
This table reports all available performance indicators, including exact score matching (SMR Exact), tolerance-based matching (SMR $\pm 1$), and granular error metrics (MAE, RMSE) for both final scores and intermediate counts (e.g., recall number, item accuracy).
}
\setlength{\tabcolsep}{15pt}
\label{tab:full_exam}
\scalebox{0.9}{
\begin{tabular}{llc}
\toprule
\textbf{Task} & \textbf{Metric} & \textbf{Value} \\
\midrule
\multicolumn{3}{l}{\textit{MoCA-SL}} \\
\midrule
\multirow{5}{*}{Picture Naming} 
 & SMR (Exact) & 97.1\% \\
 & SMR ($\pm 1$) & 100.0\% \\
 & MAE (Score) & 0.029 \\
 & RMSE (Score) & 0.171 \\
\grayrule

\multirow{6}{*}{Digit Span} 
 & SMR (Exact) & 98.5\% \\
 & SMR ($\pm 1$) & 100.0\% \\
 & MAE (Score) & 0.015 \\
 & RMSE (Score) & 0.123 \\
 & Forward Accuracy & 100.0\% \\
 & Backward Accuracy & 98.5\% \\
\grayrule

\multirow{7}{*}{Serial 7 Subtraction} 
 & SMR (Exact) & 82.4\% \\
 & SMR ($\pm 1$) & 98.5\% \\
 & MAE (Score) & 0.191 \\
 & RMSE (Score) & 0.470 \\
 & MAE (Count Correct) & 0.368 \\
 & RMSE (Count Correct) & 0.813 \\
\grayrule

\multirow{6}{*}{Sentence Repetition} 
 & SMR (Exact) & 89.7\% \\
 & SMR ($\pm 1$) & 100.0\% \\
 & MAE (Score) & 0.103 \\
 & RMSE (Score) & 0.321 \\
 & Q-1 Accuracy & 89.7\% \\
 & Q-2 Accuracy & 100.0\% \\
\grayrule

\multirow{7}{*}{Animal Fluency} 
 & SMR (Exact) & 98.5\% \\
 & SMR ($\pm 1$) & 100.0\% \\
 & MAE (Score) & 0.015 \\
 & RMSE (Score) & 0.121 \\
 & MAE (Count) & 0.544 \\
\grayrule

\multirow{6}{*}{Abstraction} 
 & SMR (Exact) & 82.4\% \\
 & SMR ($\pm 1$) & 100.0\% \\
 & MAE (Score) & 0.176 \\
 & RMSE (Score) & 0.420 \\
 & Q-1 Accuracy & 86.8\% \\
 & Q-2 Accuracy & 86.8\% \\

\midrule
\multicolumn{3}{l}{\textit{HKLLT}} \\
\midrule

\multirow{7}{*}{Trial-4} 
 & MAE (Recall N) & 0.059 \\
 & RMSE (Recall N) & 0.243 \\
 & SMR (Recall N) & 94.1\% \\
 & SMR (Recall N, $\pm 1$) & 100.0\% \\
 & MAE (Z-Score) & 0.084 \\
 & RMSE (Z-Score) & 0.380 \\
 & SMR (Z-Score) & 88.2\% \\
\grayrule

\multirow{7}{*}{Trial-5} 
 & MAE (Recall N) & 0.074 \\
 & RMSE (Recall N) & 0.271 \\
 & SMR (Recall N) & 92.6\% \\
 & SMR (Recall N, $\pm 1$) & 100.0\% \\
 & MAE (Z-Score) & 0.109 \\
 & RMSE (Z-Score) & 0.453 \\
 & SMR (Z-Score) & 86.8\% \\

\bottomrule
\end{tabular}}
\end{table}
Table~\ref{tab:full_exam} extends the main evaluation (Table 2) with a comprehensive breakdown of performance indicators, including exact score matching (SMR Exact), tolerance-based matching (SMR $\pm$1), and granular error metrics (MAE, RMSE) for both final scores and intermediate counts.


\section{Prompts and System Output}
\label{appendix:prompts}

\subsection{Prompt for Examiner Agent}
\label{appendix:prompts:examiner}

All Examiner agents follow a standardized four-component template structure.
Each prompt comprises: (1) Task Introduction defining the agent's clinical role and assessment objective, (2) Guidelines specifying processing rules and edge cases, (3) Output Format detailing the required response structure, and (4) Examples providing demonstrations for consistent behavior.

We provide two representative examples of Examiner prompts below. Figure~\ref{fig:abs-prompt} presents the prompt for the MoCA Abstraction task, which requires the agent to extract the participant's responses to two similarity questions and judge whether each response reflects abstract categorical reasoning (e.g., transportation'' for train-bicycle) versus concrete associations (e.g., they have wheels''). Figure~\ref{fig:animal-naming-prompt}  presents the prompt for the MoCA Animal Naming Fluency task, which instructs the agent to extract all valid animal names from the participant's verbal response, handle Cantonese colloquial expressions, deduplicate semantically equivalent items, and invoke the \texttt{list\_length()} function to obtain the final count.

\begin{figure*}[htbp] 
    \centering
    \begin{tcolorbox}[colback=gray!5!white, colframe=gray!50!black, title=\textbf{Prompt Example: MoCA Abstraction (Bilingual Contrast)}]
        
        \begin{CJK*}{UTF8}{bsmi}
            \scriptsize 
            
            \begin{minipage}[t]{0.48\linewidth}
                - Original Prompt -\\
                \vspace{0.2cm}
                \textbf{[Instruction]:} \\
                \textbf{\#\# 任務介紹} \\
                你是一位醫學助理，評估MoCA抽象概唸任務。任務是從轉錄中提取受試者對兩個問題的回答（Q1:火車-單車相似處, Q2:手表-直尺相似處），並判斷是否正確。
                
                \vspace{0.1cm}
                \textbf{\#\# 注意事項}\\
                - Q1和Q2之間有 \texttt{\textless|question-change|\textgreater} 分隔\\
                - Q1正確答案：交通工具、運輸工具、代步工具等（錯誤：有輪子、會動、沒區別、唔相似）\\
                - Q2正確答案：度量工具、測量工具、計量器具等（錯誤：有數字、可以看時間、不知道、諗唔到）\\
                - 轉錄爲粵語口語

                \vspace{0.1cm}
                \textbf{\#\# 格式要求} \\
                Q1回答: \textless 提取概唸\textgreater{} \quad Q1判斷: \texttt{\textless true/false\textgreater} \\
                Q2回答: \textless 提取概唸\textgreater{} \quad \quad Q2判斷: \texttt{\textless true/false\textgreater} \\
                \texttt{\{ "Q1": \{"response": ["..."], "is\_correct": \textless true/false\textgreater\}, "Q2": \{"response": ["..."], "is\_correct": \textless true/false\textgreater\} \}}

                \vspace{0.1cm}
                \textbf{\#\# 範例 1} \\
                轉錄: \texttt{"單車同火車, 都繫交通嘅工具, \textless|question-change|\textgreater, 手表, 間尺, 大家都繫, 計數目嘅..."} \\
                Q1回答: 交通工具 \quad Q1判斷: true \\
                Q2回答: 計數目 \quad \quad Q2判斷: false \\
                \texttt{\{ "Q1": \{"response": ["交通工具"], "is\_correct": true\}, "Q2": \{"response": ["計數目"], "is\_correct": false\} \}}
                
                \vspace{0.1cm}
                \rule{\linewidth}{0.4pt} 
                \vspace{0.1cm}
                
                \textbf{[User Input]:} \\
                \textbf{\#\# 轉錄} \\
                請按格式要求輸出Q1、Q2回答和判斷，並輸出JSON結果。\\
                轉錄: \texttt{"誒 交通工具囉, 運輸工具囉, \textless|question-change|\textgreater, 都繫量度一啲嘅嘢囉, 一個時間一個長度"}
                
                \vspace{0.2cm}
                \textbf{[Model Output (Intermediate)]:} \\
                Q1回答: 交通工具 \quad Q1判斷: true \\
                Q2回答: 量度一啲嘅嘢 \quad Q2判斷: false \\
                \texttt{\{ "Q1": \{"response": ["交通工具"], "is\_correct": true\}, "Q2": \{"response": ["量度一啲嘅嘢"], "is\_correct": false\} \}}
                
                \vspace{0.2cm}
                \textbf{[Verifier]:} \\
                Q2判斷錯誤（假陰性）：回答 "量度一啲嘅嘢" 包含度量概唸，應該判爲正確 \\
                - 模型判斷: is\_correct = false \\
                - 正確判斷: is\_correct = true \\
                修正建議：將 Q2.is\_correct 改成 true
                
                \vspace{0.2cm}
                \textbf{[Final Model Output]:} \\
                \texttt{\{
                    "Q1": \{"response": ["交通工具"], "is\_correct": true\},
                    "Q2": \{"response": ["量度一啲嘅嘢"], "is\_correct": true\}
                \}}

                \vspace{0.2cm}
                \textbf{[Verifier]:} \\
                所有回答都在轉錄中找到且判斷正確: \\
                - Q1回答 ''交通工具'' (is\_correct=True) \\
                - Q2回答 ''量度一啲嘅嘢'' (is\_correct=True)\\
            \end{minipage}%
            \hfill\vrule\hfill 
            \begin{minipage}[t]{0.48\linewidth}
                - Translated English Prompt -\\
                \vspace{0.2cm}
                \textbf{[Instruction]:} \\
                \textbf{\#\# Task Description} \\
                You are a medical assistant evaluating the MoCA Abstraction task. Extract the subject''s answers to two questions (Q1: Train-Bicycle similarity, Q2: Watch-Ruler similarity) from the transcript and determine correctness.
                
                \vspace{0.1cm}
                \textbf{\#\# Notes}\\
                - Q1 and Q2 are separated by \texttt{\textless|question-change|\textgreater}.\\
                - Q1 Correct: Transportation, Vehicle, Commute tools, etc. (Wrong: have wheels, move, no difference, not similar).\\
                - Q2 Correct: Measuring tools, Measurement instruments, Gauges, etc. (Wrong: have numbers, tells time, don''t know, can''t think of it).\\
                - Transcript is in Cantonese colloquial.

                \vspace{0.1cm}
                \textbf{\#\# Format Requirements} \\
                Q1 Resp: \textless Extracted Concept\textgreater{} \quad Q1 Judge: \texttt{\textless true/false\textgreater} \\
                Q2 Resp: \textless Extracted Concept\textgreater{} \quad Q2 Judge: \texttt{\textless true/false\textgreater} \\
                \texttt{\{ "Q1": \{"response": ["..."], "is\_correct": \textless true/false\textgreater\}, "Q2": \{"response": ["..."], "is\_correct": \textless true/false\textgreater\} \}}

                \vspace{0.1cm}
                \textbf{\#\# Example 1} \\
                Transcript: \texttt{"Bicycle and train, both are transport tools, \textless|question-change|\textgreater, watch, ruler, everyone is, counting numbers..."} \\
                Q1 Resp: Transport tools \quad Q1 Judge: true \\
                Q2 Resp: Counting numbers \quad Q2 Judge: false \\
                \texttt{\{ "Q1": \{"response": ["Transport tools"], "is\_correct": true\}, "Q2": \{"response": ["Counting numbers"], "is\_correct": false\} \}}
                
                \vspace{0.1cm}
                \rule{\linewidth}{0.4pt} 
                \vspace{0.1cm}
                
                \textbf{[User Input]:} \\
                \textbf{\#\# Transcript} \\
                Please output Q1/Q2 response and judgment in JSON format.\\
                Transcript: \texttt{"Eh transport tools lo, transportation tools lo, \textless|question-change|\textgreater, both measure things lo, one time one length"}
                
                \vspace{0.2cm}
                \textbf{[Model Output (Intermediate)]:} \\
                Q1 Resp: transport tools \quad Q1 Judge: true \\
                Q2 Resp: measure things \quad Q2 Judge: false \\
                \texttt{\{ "Q1": \{"response": ["transport tools"], "is\_correct": true\}, "Q2": \{"response": ["measure things"], "is\_correct": false\} \}}
                
                \vspace{0.2cm}
                \textbf{[Verifier]:} \\
                Q2 Judgment Error (False Negative): Response "measure things" contains measurement concept, should be correct \\
                - Model: is\_correct = false \\
                - Ground Truth: is\_correct = true \\
                Correction: Change Q2.is\_correct to true.
                
                \vspace{0.2cm}
                \textbf{[Final Model Output]:} \\
                \texttt{\{
                    "Q1": \{"response": ["transport tools"], "is\_correct": true\},
                    "Q2": \{"response": ["measure things"], "is\_correct": true\}
                \}}

                \vspace{0.2cm}
                \textbf{[Verifier]:} \\
                All responses found in transcript and correctly judged: \\
                - Q1 response ''transport tools'' (is\_correct=True) \\
                - Q2 response ''measure things'' (is\_correct=True)\\
            \end{minipage}
            
        \end{CJK*}
    \end{tcolorbox}
    \caption{An example of the Examiner's prompt with a verifier step for MoCA Abstraction task. The left column shows the original Cantonese prompt, and the right column shows the English translation.}
    \label{fig:abs-prompt}
\end{figure*}
\begin{figure*}[htbp] 
    \centering
    \begin{tcolorbox}[colback=gray!5!white, colframe=gray!50!black, title=\textbf{Prompt Example: MoCA Animal Naming Fluency (Bilingual Contrast)}]
        
        \begin{CJK*}{UTF8}{bsmi}
            \scriptsize 
            
            \begin{minipage}[t]{0.48\linewidth}
                - Original Prompt -\\
                \vspace{0.2cm}
                \textbf{[Instruction]:} \\
                \textbf{\#\# 任務介紹} \\
                你是一位醫學助理，評估MoCA動物命名流暢性任務。任務是從轉錄中提取受試者説出的所有動物名稱，然後調用list\_length()函數計數。
                
                \vspace{0.1cm}
                \textbf{\#\# 注意事項}\\
                - 提取所有動物名稱，按順序記錄\\
                - 重複的動物隻計算一次\\
                - 轉錄爲粵語口語，需理解粵語(如：馬騮/馬餾=猴子)

                \vspace{0.1cm}
                \textbf{\#\# 格式要求} \\
                轉錄: \textless 原文片段\textgreater \\
                動物列表: [動物1, 動物2, 動物3, ...] \\
                \texttt{\textless tool\_call\textgreater} \\
                \texttt{\{"name": "list\_length", "arguments": \{"list": ["動物1", "動物2", ...]\}\}} \\
                \texttt{\textless /tool\_call\textgreater}

                \vspace{0.1cm}
                \textbf{\#\# 範例} \\
                轉錄: \texttt{"嗯, 獅子, 老虎, 豹, 駱駝, 大笨象, 熊, 熊同埋狐狸, 狼, 猴子, 誒 野貓, 恐龍, 嗯"} \\
                動物列表: [獅子, 老虎, 豹, 駱駝, 大笨象, 熊, 狐狸, 狼, 猴子, 野貓, 恐龍] \\
                \texttt{\textless tool\_call\textgreater} \\
                \texttt{\{"name": "list\_length", "arguments": \{"list": ["獅子", "老虎", "豹", "駱駝", "大笨象", "熊", "狐狸", "狼", "猴子", "野貓", "恐龍"]\}\}} \\
                \texttt{\textless /tool\_call\textgreater}
                
                \vspace{0.1cm}
                \rule{\linewidth}{0.4pt} 
                \vspace{0.1cm}
                
                \textbf{[User Input]:} \\
                \textbf{\#\# 轉錄} \\
                請按格式要求輸出動物列表並調用函數。\\
                轉錄: \texttt{"哦, 越多越好, 誒即繫即繫水上又得, 即繫繫咪, 哦噉得啦得啦, 鼠 老鼠 牛誒牛貓狗, 誒 羊, 駱, 我繫飛誒飛蛾, 長嗯長頸鹿, 大笨象, 啊 馬, 羊山羊, 天誒嗰隻誒, 有誒水水, 嗰隻叫龜啊水 龜, 龜同埋, 誒, 啊"}
                
                \vspace{0.2cm}
                \textbf{[Model Output]:} \\
                動物列表: [老鼠, 牛, 貓, 狗, 羊, 駱駝, 飛蛾, 長頸鹿, 大笨象, 馬, 山羊, 龜] \\
                \texttt{\textless tool\_call\textgreater} \\
                \texttt{\{"name": "list\_length", "arguments": \{"list": ["老鼠", "牛", "貓", "狗", "羊", "熊", "駱駝", "飛蛾", "長頸鹿", "大笨象", "馬", "山羊", "龜"]\}\}} \\
                \texttt{\textless /tool\_call\textgreater}
                
                \vspace{0.2cm}
                \textbf{[Verifier]:} \\
                Pass, 所有 12 個動物都在轉錄中找到。
                
            \end{minipage}%
            \hfill\vrule\hfill 
            \begin{minipage}[t]{0.48\linewidth}
                - Translated English Prompt -\\
                \vspace{0.2cm}
                \textbf{[Instruction]:} \\
                \textbf{\#\# Task Description} \\
                You are a medical assistant evaluating the MoCA Animal Naming Fluency task. Extract all animal names mentioned by the subject from the transcript, then call the list\_length() function to count them.
                
                \vspace{0.1cm}
                \textbf{\#\# Notes}\\
                - Extract all animal names in order\\
                - Repeated animals are counted only once\\
                - Transcript is in Cantonese colloquial, need to understand Cantonese (e.g., 馬騮/馬餾 = monkey)

                \vspace{0.1cm}
                \textbf{\#\# Format Requirements} \\
                Transcript: \textless original text\textgreater \\
                Animal list: [Animal1, Animal2, Animal3, ...] \\
                \texttt{\textless tool\_call\textgreater} \\
                \texttt{\{"name": "list\_length", "arguments": \{"list": ["Animal1", "Animal2", ...]\}\}} \\
                \texttt{\textless /tool\_call\textgreater}

                \vspace{0.1cm}
                \textbf{\#\# Example} \\
                Transcript: \texttt{"Um, lion, tiger, leopard, camel, elephant, bear, bear and fox, wolf, monkey, eh wild cat, dinosaur, um"} \\
                Animal list: [Lion, Tiger, Leopard, Camel, Elephant, Bear, Fox, Wolf, Monkey, Wild cat, Dinosaur] \\
                \texttt{\textless tool\_call\textgreater} \\
                \texttt{\{"name": "list\_length", "arguments": \{"list": ["Lion", "Tiger", "Leopard", "Camel", "Elephant", "Bear", "Fox", "Wolf", "Monkey", "Wild cat", "Dinosaur"]\}\}} \\
                \texttt{\textless /tool\_call\textgreater}
                
                \vspace{0.1cm}
                \rule{\linewidth}{0.4pt} 
                \vspace{0.1cm}
                
                \textbf{[User Input]:} \\
                \textbf{\#\# Transcript} \\
                Please output the animal list and call the function as required.\\
                Transcript: \texttt{"Oh, the more the better, eh that is on water also works, that is right, oh then ok ok, mouse rat cow eh cow cat dog, eh sheep, cam, I am fly eh moth, long um giraffe, elephant, ah horse, sheep goat, sky eh that one eh, have eh water water, that one called turtle ah water turtle, turtle and, eh, ah"}
                
                \vspace{0.2cm}
                \textbf{[Model Output]:} \\
                Animal list: [Rat, Cow, Cat, Dog, Sheep, Camel, Moth, Giraffe, Elephant, Horse, Goat, Turtle] \\
                \texttt{\textless tool\_call\textgreater} \\
                \texttt{\{"name": "list\_length", "arguments": \{"list": ["Rat", "Cow", "Cat", "Dog", "Sheep", "Bear", "Camel", "Moth", "Giraffe", "Elephant", "Horse", "Goat", "Turtle"]\}\}} \\
                \texttt{\textless /tool\_call\textgreater}
                
                \vspace{0.2cm}
                \textbf{[Verifier]:} \\
                Pass, all 12 animals found in transcript.
                
            \end{minipage}
            
        \end{CJK*}
    \end{tcolorbox}
    \caption{
    An example of the Examiner's prompt with a verifier step for MoCA Animal Fluency task. The left column shows the original Cantonese prompt, and the right column shows the English translation.}
    
    \label{fig:animal-naming-prompt}
\end{figure*}

\subsection{Meta Analyst}
\label{appendix:prompts:analyst}

The Meta Analyst agent transforms verified scoring primitives into interpretable cognitive profile reports. 
Its instruction prompt comprises two components: (1) \textbf{Clinical Protocol Context} (Figure~\ref{fig:knowledge}), 
which provides the evaluation framework including normal ranges, impairment thresholds, and clinical significance 
for each HKLLT and MoCA subtest; and (2) \textbf{Output Requirements} (Figure~\ref{fig:cognitive-report-prompt}, upper panel), 
which specifies the report format—assessments organized by four cognitive domains (memory, executive function, 
attention \& working memory, language), each containing status judgment, supporting evidence, and clinical interpretation.

Figure~\ref{fig:cognitive-report-prompt} presents an example input for a 75-year-old male with 6 years of education, 
showing HKLLT z-scores in the mild-to-moderate impairment range and mixed MoCA subtest performance. 
Figure~\ref{fig:profile1} displays the corresponding model output, where the Meta Analyst identifies moderate memory 
impairment based on delayed recall z-scores (-1.5 to -2.0 range), mild executive and language dysfunction, 
while correctly recognizing preserved attention and working memory—culminating in a HIGH risk assessment 
with recommendation for further clinical evaluation.

\begin{figure*}[htbp]
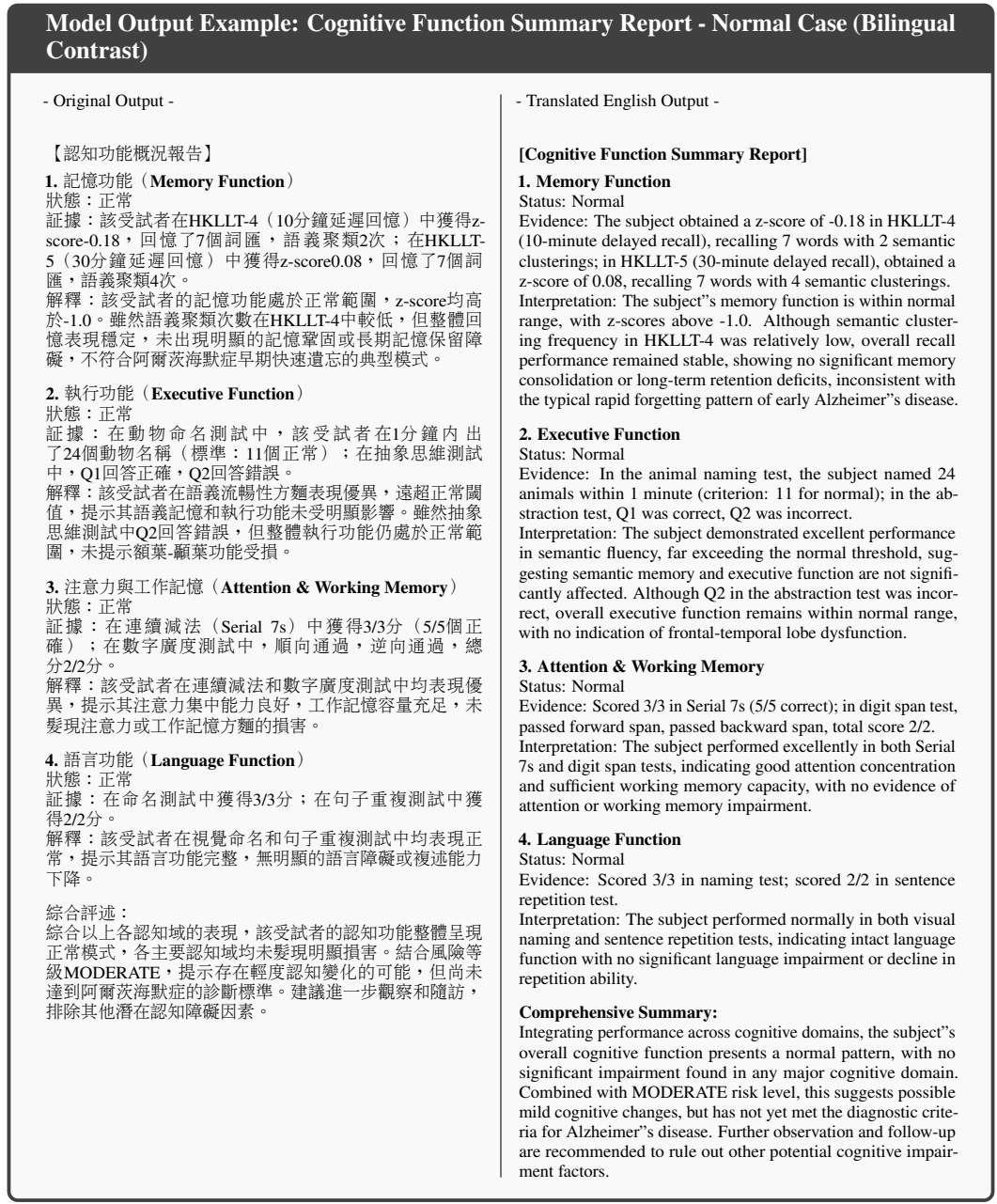
 
    \centering
    \begin{tcolorbox}[colback=gray!5!white, colframe=gray!50!black, title=\textbf{Model Output Example: Cognitive Function Summary Report - Normal Case (Bilingual Contrast)}]
        
        \begin{CJK*}{UTF8}{bsmi}
            \scriptsize 
            
            \begin{minipage}[t]{0.48\linewidth}
                - Original Output -\\
                \vspace{0.2cm}
                
                \textbf{【認知功能概況報告】}
                
                \vspace{0.1cm}
                \textbf{1. 記憶功能（Memory Function）}\\
                狀態：正常\\
                証據：該受試者在HKLLT-4（10分鐘延遲回憶）中獲得z-score爲-0.18，回憶了7個詞匯，語義聚類2次；在HKLLT-5（30分鐘延遲回憶）中獲得z-score爲0.08，回憶了7個詞匯，語義聚類4次。\\
                解釋：該受試者的記憶功能處於正常範圍，z-score均高於-1.0。雖然語義聚類次數在HKLLT-4中較低，但整體回憶表現穩定，未出現明顯的記憶鞏固或長期記憶保留障礙，不符合阿爾茨海默症早期快速遺忘的典型模式。
                
                \vspace{0.2cm}
                \textbf{2. 執行功能（Executive Function）}\\
                狀態：正常\\
                証據：在動物命名測試中，該受試者在1分鐘內説出了24個動物名稱（標準：≥11個爲正常）；在抽象思維測試中，Q1回答正確，Q2回答錯誤。\\
                解釋：該受試者在語義流暢性方麵表現優異，遠超正常閾值，提示其語義記憶和執行功能未受明顯影響。雖然抽象思維測試中Q2回答錯誤，但整體執行功能仍處於正常範圍，未提示額葉-顳葉功能受損。
                
                \vspace{0.2cm}
                \textbf{3. 注意力與工作記憶（Attention \& Working Memory）}\\
                狀態：正常\\
                証據：在連續減法（Serial 7s）中獲得3/3分（5/5個正確）；在數字廣度測試中，順向通過，逆向通過，總分2/2分。\\
                解釋：該受試者在連續減法和數字廣度測試中均表現優異，提示其注意力集中能力良好，工作記憶容量充足，未髮現注意力或工作記憶方麵的損害。
                
                \vspace{0.2cm}
                \textbf{4. 語言功能（Language Function）}\\
                狀態：正常\\
                証據：在命名測試中獲得3/3分；在句子重複測試中獲得2/2分。\\
                解釋：該受試者在視覺命名和句子重複測試中均表現正常，提示其語言功能完整，無明顯的語言障礙或複述能力下降。
                
                \vspace{0.2cm}
                \textbf{綜合評述：}\\
                綜合以上各認知域的表現，該受試者的認知功能整體呈現正常模式，各主要認知域均未髮現明顯損害。結合風險等級爲MODERATE，提示存在輕度認知變化的可能，但尚未達到阿爾茨海默症的診斷標準。建議進一步觀察和隨訪，排除其他潛在認知障礙因素。
                
            \end{minipage}%
            \hfill\vrule\hfill 
            \begin{minipage}[t]{0.48\linewidth}
                - Translated English Output -\\
                \vspace{0.2cm}
                
                \textbf{[Cognitive Function Summary Report]}
                
                \vspace{0.1cm}
                \textbf{1. Memory Function}\\
                Status: Normal\\
                Evidence: The subject obtained a z-score of -0.18 in HKLLT-4 (10-minute delayed recall), recalling 7 words with 2 semantic clusterings; in HKLLT-5 (30-minute delayed recall), obtained a z-score of 0.08, recalling 7 words with 4 semantic clusterings.\\
                Interpretation: The subject''s memory function is within normal range, with z-scores above -1.0. Although semantic clustering frequency in HKLLT-4 was relatively low, overall recall performance remained stable, showing no significant memory consolidation or long-term retention deficits, inconsistent with the typical rapid forgetting pattern of early Alzheimer''s disease.
                
                \vspace{0.2cm}
                \textbf{2. Executive Function}\\
                Status: Normal\\
                Evidence: In the animal naming test, the subject named 24 animals within 1 minute (criterion: ≥11 for normal); in the abstraction test, Q1 was correct, Q2 was incorrect.\\
                Interpretation: The subject demonstrated excellent performance in semantic fluency, far exceeding the normal threshold, suggesting semantic memory and executive function are not significantly affected. Although Q2 in the abstraction test was incorrect, overall executive function remains within normal range, with no indication of frontal-temporal lobe dysfunction.
                
                \vspace{0.2cm}
                \textbf{3. Attention \& Working Memory}\\
                Status: Normal\\
                Evidence: Scored 3/3 in Serial 7s (5/5 correct); in digit span test, passed forward span, passed backward span, total score 2/2.\\
                Interpretation: The subject performed excellently in both Serial 7s and digit span tests, indicating good attention concentration and sufficient working memory capacity, with no evidence of attention or working memory impairment.
                
                \vspace{0.2cm}
                \textbf{4. Language Function}\\
                Status: Normal\\
                Evidence: Scored 3/3 in naming test; scored 2/2 in sentence repetition test.\\
                Interpretation: The subject performed normally in both visual naming and sentence repetition tests, indicating intact language function with no significant language impairment or decline in repetition ability.
                
                \vspace{0.2cm}
                \textbf{Comprehensive Summary:}\\
                Integrating performance across cognitive domains, the subject''s overall cognitive function presents a normal pattern, with no significant impairment found in any major cognitive domain. Combined with MODERATE risk level, this suggests possible mild cognitive changes, but has not yet met the diagnostic criteria for Alzheimer''s disease. Further observation and follow-up are recommended to rule out other potential cognitive impairment factors.
                
            \end{minipage}
            
        \end{CJK*}
    \end{tcolorbox}
    \caption{An example of the model-generated cognitive function summary report for a normal case. The left column shows the original Chinese output, and the right column shows the English translation.}
    \label{fig:profile2}
\end{figure*}

Figure~\ref{fig:profile2} illustrates another output for a cognitively normal participant. 
Despite one incorrect abstraction item, the agent appropriately concludes that overall executive function 
remains within normal range, demonstrating the framework's ability to contextualize isolated errors 
within the broader performance pattern rather than over-interpreting individual task failures.
\begin{figure*}[htbp] 
    \centering
    \begin{tcolorbox}[colback=gray!5!white, colframe=gray!50!black, title=\textbf{Clinical Protocol Context (Bilingual Contrast)}]
        
        \begin{CJK*}{UTF8}{bsmi}
            \scriptsize 
            
            \begin{minipage}[t]{0.48\linewidth}
                - Original Version -\\
                \vspace{0.2cm}
                
                \textbf{\# 評估框架}
                
                \vspace{0.2cm}
                \textbf{\#\# 核心評估指標及臨床意義}
                
                \vspace{0.2cm}
                \textbf{\#\#\# 1. HKLLT記憶測試（Hong Kong List Learning Test）}

                \vspace{0.1cm}
                \textbf{正常範圍與損害分級：}
                \begin{itemize}[leftmargin=*, nosep, label=-]
                    \item 正常範圍: z-score $>$ -1.0
                    \item 輕度損害: z-score -1.0 至 -1.5
                    \item 中度損害: z-score -1.5 至 -2.0
                    \item 重度損害: z-score $<$ -2.0
                \end{itemize}

                \vspace{0.1cm}
                \textbf{臨床意義：}\\
                情景記憶是阿爾茨海默症最早受損的認知域。HKLLT z-score是核心診斷指標。
                \begin{itemize}[leftmargin=*, nosep, label=-]
                    \item Trial 4（10分鐘延遲回憶）: 評估短期記憶鞏固
                    \item Trial 5（30分鐘延遲回憶）: 評估長期記憶保留
                    \item 語義聚類次數: 反映記憶組織策略的有效性
                \end{itemize}
                
                \vspace{0.2cm}
                \textbf{\#\#\# 2. MoCA認知評估子項}

                \vspace{0.1cm}
                \textbf{2.1 動物命名測試（Semantic Fluency）}
                \begin{itemize}[leftmargin=*, nosep, label=-]
                    \item 正常: $≥$11個動物名稱/分鐘
                    \item 異常: $<$11個
                    \item 臨床意義: 評估語義流暢性和執行功能，對額葉-顳葉功能敏感
                \end{itemize}

                \vspace{0.1cm}
                \textbf{2.2 連續減法（Serial 7s）}
                \begin{itemize}[leftmargin=*, nosep, label=-]
                    \item 滿分: 3分（4-5個正確）
                    \item 輕度損害: 2分（2-3個正確）
                    \item 重度損害: 0-1分
                    \item 臨床意義: 評估注意力、工作記憶和計算能力
                \end{itemize}

                \vspace{0.1cm}
                \textbf{2.3 命名測試（Visual Naming）}
                \begin{itemize}[leftmargin=*, nosep, label=-]
                    \item 滿分: 3分（獅子、犀牛、駱駝）
                    \item 臨床意義: 評估視覺命名能力，對語義記憶敏感
                \end{itemize}

                \vspace{0.1cm}
                \textbf{2.4 句子重複（Sentence Repetition）}
                \begin{itemize}[leftmargin=*, nosep, label=-]
                    \item 滿分: 2分
                    \item 臨床意義: 評估語言功能和工作記憶
                \end{itemize}

                \vspace{0.1cm}
                \textbf{2.5 數字廣度（Digit Span）}
                \begin{itemize}[leftmargin=*, nosep, label=-]
                    \item 滿分: 2分（順向+逆向）
                    \item 臨床意義: 評估注意力和工作記憶容量
                \end{itemize}

                \vspace{0.1cm}
                \textbf{2.6 抽象思維（Abstraction）}
                \begin{itemize}[leftmargin=*, nosep, label=-]
                    \item 滿分: 2分
                    \item 臨床意義: 評估執行功能和概唸推理能力
                \end{itemize}
                
            \end{minipage}%
            \hfill\vrule\hfill 
            \begin{minipage}[t]{0.48\linewidth}
                - Translated English Version -\\
                \vspace{0.2cm}
                
                \textbf{\# Evaluation Framework}
                
                \vspace{0.2cm}
                \textbf{\#\# Core Assessment Indicators and Clinical Significance}
                
                \vspace{0.2cm}
                \textbf{\#\#\# 1. HKLLT Memory Test (Hong Kong List Learning Test)}

                \vspace{0.1cm}
                \textbf{Normal Range and Impairment Levels:}
                \begin{itemize}[leftmargin=*, nosep, label=-]
                    \item Normal range: z-score $>$ -1.0
                    \item Mild impairment: z-score -1.0 to -1.5
                    \item Moderate impairment: z-score -1.5 to -2.0
                    \item Severe impairment: z-score $>$ -2.0
                \end{itemize}

                \vspace{0.1cm}
                \textbf{Clinical Significance:}\\
                Episodic memory is the earliest impaired cognitive domain in Alzheimer''s disease. HKLLT z-score is a core diagnostic indicator.
                \begin{itemize}[leftmargin=*, nosep, label=-]
                    \item Trial 4 (10-min delayed recall): Assesses short-term memory consolidation
                    \item Trial 5 (30-min delayed recall): Assesses long-term memory retention
                    \item Semantic clustering count: Reflects effectiveness of memory organization strategies
                \end{itemize}
                
                \vspace{0.2cm}
                \textbf{\#\#\# 2. MoCA Cognitive Assessment Subtests}

                \vspace{0.1cm}
                \textbf{2.1 Animal Naming Test (Semantic Fluency)}
                \begin{itemize}[leftmargin=*, nosep, label=-]
                    \item Normal: ≥11 animal names/minute
                    \item Abnormal: <11
                    \item Clinical significance: Assesses semantic fluency and executive function, sensitive to frontal-temporal lobe function
                \end{itemize}

                \vspace{0.1cm}
                \textbf{2.2 Serial Subtraction (Serial 7s)}
                \begin{itemize}[leftmargin=*, nosep, label=-]
                    \item Full score: 3 points (4-5 correct)
                    \item Mild impairment: 2 points (2-3 correct)
                    \item Severe impairment: 0-1 points
                    \item Clinical significance: Assesses attention, working memory, and calculation ability
                \end{itemize}

                \vspace{0.1cm}
                \textbf{2.3 Naming Test (Visual Naming)}
                \begin{itemize}[leftmargin=*, nosep, label=-]
                    \item Full score: 3 points (Lion, Rhino, Camel)
                    \item Clinical significance: Assesses visual naming ability, sensitive to semantic memory
                \end{itemize}

                \vspace{0.1cm}
                \textbf{2.4 Sentence Repetition}
                \begin{itemize}[leftmargin=*, nosep, label=-]
                    \item Full score: 2 points
                    \item Clinical significance: Assesses language function and working memory
                \end{itemize}

                \vspace{0.1cm}
                \textbf{2.5 Digit Span}
                \begin{itemize}[leftmargin=*, nosep, label=-]
                    \item Full score: 2 points (Forward + Backward)
                    \item Clinical significance: Assesses attention and working memory capacity
                \end{itemize}

                \vspace{0.1cm}
                \textbf{2.6 Abstraction}
                \begin{itemize}[leftmargin=*, nosep, label=-]
                    \item Full score: 2 points
                    \item Clinical significance: Assesses executive function and conceptual reasoning ability
                \end{itemize}
                
            \end{minipage}
            
        \end{CJK*}
    \end{tcolorbox}
    \caption{Clinical Protocol Context is a section in Meta Analyst's instruction prompt, showing core assessment indicators and clinical significance. The left column shows the original Chinese version, and the right column shows the English translation.}
    \label{fig:knowledge}
\end{figure*}
\begin{figure*}[htbp] 
    \centering
    \scalebox{0.85}{
    \begin{tcolorbox}[colback=gray!5!white, colframe=gray!50!black, title=\textbf{Prompt Example: Cognitive Profile Report Generation (Bilingual Contrast)}]
        
        \begin{CJK*}{UTF8}{bsmi}
            \scriptsize 
            
            \begin{minipage}[t]{0.48\linewidth}
                - Original Prompt -\\
                \vspace{0.2cm}
                \textbf{[Instruction]:} \\
                \textbf{\# Clinical Protocol Context:} \\
                (..省略..)
                
                \vspace{0.1cm}
                \textbf{\# 輸出要求} \\
                請以\textbf{自然語言陳述}的形式，爲該受試者撰冩一份詳細的認知功能概況報告。報告應包含以下四個認知域的評估，每個域需包含：狀態判斷、支持証據、臨床解釋。

                \vspace{0.1cm}
                \textbf{報告格式範例：}
                
                \texttt{【認知功能概況報告】}

                \vspace{0.1cm}
                \textbf{1. 記憶功能（Memory Function）}\\
                狀態：[正常/輕度損害/中度損害/重度損害]\\
                証據：例如，該受試者在HKLLT-4（10分鐘延遲回憶）中獲得z-score爲X，回憶了Y個詞匯，語義聚類Z次；在HKLLT-5（30分鐘延遲回憶）中獲得z-score爲X，回憶了Y個詞匯。\\
                解釋：[根據z-score閾值和表現模式，説明記憶功能的損害程度及其臨床意義...]

                \vspace{0.1cm}
                \textbf{2. 執行功能（Executive Function）}\\
                狀態：[正常/損害]\\
                証據：在動物命名測試中，該受試者在1分鐘內説出了X個動物名稱（標準：≥11個爲正常）；在抽象思維測試中，Q1回答[正確/錯誤]，Q2回答[正確/錯誤]。\\
                解釋：[説明語義流暢性和概唸推理能力的表現]

                \vspace{0.1cm}
                \textbf{3. 注意力與工作記憶}\\
                狀態：[正常/損害]\\
                証據：在連續減法（Serial 7s）中獲得X/3分（Y/5個正確）；在數字廣度測試中，順向[通過/未通過]，逆向[通過/未通過]，總分X/2分。\\
                解釋：[説明注意力持續性和工作記憶容量的表現]

                \vspace{0.1cm}
                \textbf{4. 語言功能（Language Function）}\\
                狀態：[正常/損害]\\
                証據：在命名測試中獲得X/3分；在句子重複測試中獲得X/2分。\\
                解釋：[説明視覺命名和語言複述能力的表現]

                \vspace{0.1cm}
                \textbf{綜合評述：}\\
                綜合以上各認知域的表現，該受試者的認知功能整體呈現[描述整體模式]...結合風險等級爲[LOW/MODERATE/HIGH/VERY\_HIGH]，[提供整體臨床印象]。
                
                \vspace{0.1cm}
                \rule{\linewidth}{0.4pt} 
                \vspace{0.1cm}
                
                \textbf{[User Input]:} \\
                \vspace{0.2cm}
                \textbf{\# 當前案例}

                \vspace{0.2cm}
                \textbf{\#\# 基本信息}\\
                - 年齡: 75歲\\
                - 性別: 男\\
                - 教育年限: 6.0年

                \vspace{0.2cm}
                \textbf{\#\# 評估結果}

                \vspace{0.2cm}
                \textbf{\#\#\# HKLLT指標}\\
                - HKLLT-4 z-score: -0.71 (回憶4個詞，語義聚類2次)\\
                - HKLLT-5 z-score: -0.83 (回憶3個詞，語義聚類1次)

                \vspace{0.2cm}
                \textbf{\#\#\# MoCA認知評估子項}\\
                - \textbf{動物命名}: 14個 (通過標準)\\
                - \textbf{連續減法}: 2/3分 (2/5個正確)\\
                - \textbf{數字廣度}: 1/2分 (順向pass，逆向fail)\\
                - \textbf{命名測試}: 3/3分\\
                - \textbf{句子重複}: 2/2分\\
                - \textbf{抽象思維}: Q1 pass，Q2 fail
                
                \vspace{0.2cm}
                \textbf{[Model Output]:} \\
                (..省略..)
                
            \end{minipage}%
            \hfill\vrule\hfill 
            \begin{minipage}[t]{0.48\linewidth}
                - Translated English Prompt -\\
                \vspace{0.2cm}
                \textbf{[Instruction]:} \\
                \textbf{\# Clinical Protocol Context:} \\
                (..omitted..)
                
                \vspace{0.1cm}
                \textbf{\# Output Requirements} \\
                Please write a detailed cognitive profile report for the subject in \textbf{natural language narrative} form. The report should include assessments of the following four cognitive domains, each containing: status judgment, supporting evidence, and clinical interpretation.

                \vspace{0.1cm}
                \textbf{Report Format Example:}

                \texttt{【Cognitive Profile Report】}

                \vspace{0.1cm}
                \textbf{1. Memory Function}\\
                Status: [Normal/Mild impairment/Moderate impairment/Severe impairment]\\
                Evidence: For example, the subject obtained a z-score of X in HKLLT-4 (10-minute delayed recall), recalled Y words, with Z semantic clusters; obtained a z-score of X in HKLLT-5 (30-minute delayed recall), recalled Y words.\\
                Interpretation: [Based on z-score thresholds and performance patterns, explain the degree of memory impairment and clinical significance...]

                \vspace{0.1cm}
                \textbf{2. Executive Function}\\
                Status: [Normal/Impaired]\\
                Evidence: In animal naming test, the subject named X animals within 1 minute (criterion: ≥11 for normal); in abstraction test, Q1 answered [correctly/incorrectly], Q2 answered [correctly/incorrectly].\\
                Interpretation: [Explain semantic fluency and conceptual reasoning performance]

                \vspace{0.1cm}
                \textbf{3. Attention \& Working Memory}\\
                Status: [Normal/Impaired]\\
                Evidence: In Serial 7s, scored X/3 points (Y/5 correct); in digit span test, forward [passed/failed], backward [passed/failed], total score X/2.\\
                Interpretation: [Explain sustained attention and working memory capacity performance]

                \vspace{0.1cm}
                \textbf{4. Language Function}\\
                Status: [Normal/Impaired]\\
                Evidence: Scored X/3 in naming test; scored X/2 in sentence repetition test.\\
                Interpretation: [Explain visual naming and verbal repetition abilities]

                \vspace{0.1cm}
                \textbf{Overall Summary:}\\
                Integrating performance across all cognitive domains, the subject''s overall cognitive function presents [describe overall pattern]...Combined with risk level [LOW/MODERATE/HIGH/VERY\_HIGH], [provide overall clinical impression].
                
                \vspace{0.1cm}
                \rule{\linewidth}{0.4pt} 
                \vspace{0.1cm}
                
                \textbf{[User Input]:} \\
                \vspace{0.2cm}
                \textbf{\# Current Case}

                \vspace{0.2cm}
                \textbf{\#\# Basic Information}\\
                - Age: 75 years\\
                - Gender: Male\\
                - Education: 6.0 years

                \vspace{0.2cm}
                \textbf{\#\# Assessment Results}

                \vspace{0.2cm}
                \textbf{\#\#\# HKLLT Metrics}\\
                - HKLLT-4 z-score: -0.71 (recalled 4 words, 2 semantic clusters)\\
                - HKLLT-5 z-score: -0.83 (recalled 3 words, 1 semantic cluster)

                \vspace{0.2cm}
                \textbf{\#\#\# MoCA Cognitive Sub-items}\\
                - \textbf{Animal naming}: 14 (passed criterion)\\
                - \textbf{Serial 7s}: 2/3 points (2/5 correct)\\
                - \textbf{Digit span}: 1/2 points (forward pass, backward fail)\\
                - \textbf{Naming test}: 3/3 points\\
                - \textbf{Sentence repetition}: 2/2 points\\
                - \textbf{Abstraction}: Q1 pass, Q2 fail
                
                \vspace{0.2cm}
                \textbf{[Model Output]:} \\
                (..omitted..)
                
            \end{minipage}
            
        \end{CJK*}
    \end{tcolorbox}}
    \caption{An example of the prompt for generating cognitive profile reports. The left column shows the original Cantonese prompt, and the right column shows the English translation.}
    \label{fig:cognitive-report-prompt}
\end{figure*}
\begin{figure*}[htbp] 
    \centering
    \begin{tcolorbox}[colback=gray!5!white, colframe=gray!50!black, title=\textbf{Model Output Example: Cognitive Function Summary Report (Bilingual Contrast)}]
        
        \begin{CJK*}{UTF8}{bsmi}
            \scriptsize 
            
            \begin{minipage}[t]{0.48\linewidth}
                - Original Output -\\
                \vspace{0.2cm}
                
                \textbf{【認知功能概況報告】}
                
                \vspace{0.1cm}
                \textbf{1. 記憶功能（Memory Function）}\\
                狀態：中度損害\\
                証據：該受試者在HKLLT-4（10分鐘延遲回憶）中獲得z-score爲-1.65，回憶了2個詞匯，語義聚類1次；在HKLLT-5（30分鐘延遲回憶）中獲得z-score爲-1.7，回憶了1個詞匯，語義聚類1次。\\
                解釋：該受試者在延遲回憶測試中表現明顯下降，z-score處於中度損害範圍（-1.5至-2.0）。回憶詞匯數量減少且語義聚類次數有限，提示情景記憶的鞏固和長期保留能力受損，符合阿爾茨海默症早期記憶功能損害的典型表現。
                
                \vspace{0.2cm}
                \textbf{2. 執行功能（Executive Function）}\\
                狀態：輕度損害\\
                証據：在動物命名測試中，該受試者在1分鐘內説出了16個動物名稱（通過標準）；在抽象思維測試中，Q1回答錯誤，Q2回答錯誤。\\
                解釋：動物命名測試表現正常，表明語義流暢性未受明顯影響。然而，抽象思維測試中兩題均錯誤，提示概唸推理和執行功能存在輕度損害，可能與額葉-顳葉功能異常有關。
                
                \vspace{0.2cm}
                \textbf{3. 注意力與工作記憶（Attention \& Working Memory）}\\
                狀態：正常\\
                証據：在連續減法（Serial 7s）中獲得3/3分（4/5個正確）；在數字廣度測試中，順向通過，逆向通過，總分2/2分。\\
                解釋：連續減法測試和數字廣度測試均表現良好，表明注意力持續性和工作記憶容量未受明顯影響，提示該受試者在執行任務時的注意力和信息處理能力保持正常。
                
                \vspace{0.2cm}
                \textbf{4. 語言功能（Language Function）}\\
                狀態：輕度損害\\
                証據：在命名測試中獲得1/3分；在句子重複測試中獲得2/2分。\\
                解釋：命名測試中僅能正確命名1個詞匯，提示視覺命名能力受損，可能與語義記憶或語言處理中樞功能異常有關。然而，句子重複測試表現正常，説明語言複述能力未受明顯影響。
                
                \vspace{0.2cm}
                \textbf{綜合評述：}\\
                綜合以上各認知域的表現，該受試者的認知功能整體呈現記憶功能中度損害、執行功能輕度損害、語言功能輕度損害，而注意力與工作記憶保持正常。結合風險等級爲HIGH，提示存在較高的阿爾茨海默症可能性，需進一步進行影像學和臨床評估以明確診斷。
                
            \end{minipage}%
            \hfill\vrule\hfill 
            \begin{minipage}[t]{0.48\linewidth}
                - Translated English Output -\\
                \vspace{0.2cm}
                
                \textbf{[Cognitive Function Summary Report]}
                
                \vspace{0.1cm}
                \textbf{1. Memory Function}\\
                Status: Moderate impairment\\
                Evidence: The subject obtained a z-score of -1.65 in HKLLT-4 (10-minute delayed recall), recalling 2 words with 1 semantic clustering; in HKLLT-5 (30-minute delayed recall), obtained a z-score of -1.7, recalling 1 word with 1 semantic clustering.\\
                Interpretation: The subject showed significant decline in delayed recall tests, with z-scores in the moderate impairment range (-1.5 to -2.0). The reduced number of recalled words and limited semantic clustering suggest impaired consolidation and long-term retention of episodic memory, consistent with typical early-stage memory dysfunction in Alzheimer''s disease.
                
                \vspace{0.2cm}
                \textbf{2. Executive Function}\\
                Status: Mild impairment\\
                Evidence: In the animal naming test, the subject named 16 animals within 1 minute (passed criterion); in the abstraction test, Q1 was incorrect, Q2 was incorrect.\\
                Interpretation: Normal performance in animal naming test indicates semantic fluency is not significantly affected. However, both questions in the abstraction test were incorrect, suggesting mild impairment in conceptual reasoning and executive function, possibly related to frontal-temporal lobe dysfunction.
                
                \vspace{0.2cm}
                \textbf{3. Attention \& Working Memory}\\
                Status: Normal\\
                Evidence: Scored 3/3 in Serial 7s (4/5 correct); in digit span test, passed forward span, passed backward span, total score 2/2.\\
                Interpretation: Good performance in both Serial 7s and digit span tests indicates sustained attention and working memory capacity are not significantly affected, suggesting the subject maintains normal attention and information processing abilities during task execution.
                
                \vspace{0.2cm}
                \textbf{4. Language Function}\\
                Status: Mild impairment\\
                Evidence: Scored 1/3 in naming test; scored 2/2 in sentence repetition test.\\
                Interpretation: Only correctly naming 1 item in the naming test suggests impaired visual naming ability, possibly related to semantic memory or language processing center dysfunction. However, normal performance in sentence repetition indicates language repetition ability is not significantly affected.
                
                \vspace{0.2cm}
                \textbf{Comprehensive Summary:}\\
                Integrating performance across cognitive domains, the subject''s overall cognitive function shows moderate memory impairment, mild executive function impairment, mild language impairment, while attention and working memory remain normal. Combined with HIGH risk level, this suggests a high probability of Alzheimer''s disease, requiring further imaging and clinical evaluation for definitive diagnosis.
                
            \end{minipage}
            
        \end{CJK*}
    \end{tcolorbox}
    \caption{An example of the model-generated cognitive function summary report. The left column shows the original Chinese output, and the right column shows the English translation.}
    \label{fig:profile1}
\end{figure*}

\subsection{Prompt for LLM-CoT baseline}
\label{appendix:prompts:llmcot}
The LLM-CoT baseline shares the same Clinical Protocol Context as the Meta Analyst 
(Figure~\ref{fig:knowledge}). Figure~\ref{fig:llm-cot} presents the 
output requirements, which specify a JSON format containing cognitive assessments 
per domain, three-step chain-of-thought reasoning, and a final diagnosis with 
confidence score. 
Unlike our framework, this baseline directly processes raw transcripts without intermediate scoring primitive extraction or verification.

\begin{figure*}[htbp] 
    \centering
    \scalebox{0.85}{
    \begin{tcolorbox}[colback=gray!5!white, colframe=gray!50!black, title=\textbf{Prompt Example: LLM-CoT Baseline (Bilingual Contrast)}]
        
        \begin{CJK*}{UTF8}{bsmi}
            \tiny 
            
            \begin{minipage}[t]{0.48\linewidth}
                - Original Chinese Prompt -\\
                \textbf{[Instruction]:} \\
                \textbf{\#\# 臨床協議背景} \\
                ...
                
                \vspace{0.1cm}
                \textbf{\#\# 輸出要求}\\
                請嚴格按照以下JSON格式輸出診斷結果：
                
                \vspace{0.1cm}
                \texttt{\{\\
                \quad "cognitive\_assessment": \{\\
                \quad\quad "memory\_function": \{"status": "正常/輕度損害/中度損害/重度損害",\\
                \quad\quad\quad "evidence": ["HKLLT-4表現: ...", "HKLLT-5表現: ..."],\\
                \quad\quad\quad "interpretation": "簡要解釋"\},\\
                \quad\quad "executive\_function": \{"status": "正常/損害", ..."\},\\
                \quad\quad "attention\_working\_memory": \{"status": "正常/損害", ..."\},\\
                \quad\quad "language\_function": \{"status": "正常/損害", ..."\}\\
                \quad \},\\
                \quad "risk\_analysis": \{"main\_findings": [...], "risk\_factors": [...], "protective\_factors": [...]\},\\
                \quad "step\_by\_step\_reasoning": \{\\
                \quad\quad "step\_1": \{"focus": "記憶功能分析", "observation": "...", "conclusion": "..."\},\\
                \quad\quad "step\_2": \{"focus": "多認知域綜合分析", ..."\},\\
                \quad\quad "step\_3": \{"focus": "人口統計學因素與風險整合", ..."\}\\
                \quad \},\\
                \quad "diagnosis": \{"reasoning": "...", "confidence": 0.0-1.0, "is\_alzheimer\_likely": true/false\}\\
                \}}
                
                \vspace{0.1cm}
                \textbf{注意事項：}\\
                1. step\_by\_step\_reasoning必須按照cognitive\_assessment結果逐步推理\\要明確引用前麵3個step\\
                3. 保持客觀中立，隻有當多個認知域損害時才判斷爲AD可能
                
                \vspace{0.1cm}
                \rule{\linewidth}{0.4pt}
                \vspace{0.1cm}
                
                \textbf{[User Input]:} \\
                \textbf{\#\# 當前案例}\\
                \textbf{受試者信息}\\
                ID: TBS0134, 年齡: 70歲, 性別: 女, 教育年限: 0.0年
                
                \vspace{0.1cm}
                \textbf{認知測試轉錄}\\
                \textbf{HKLLT-4 (10分鐘延遲):} 
                完全唔記得囉，繫啊，完全唔記得，一個都記唔到，\\
                \textbf{HKLLT-5 (30分鐘延遲):} 啊，唔記得啦，繫啊，繫啊，\\
                \textbf{MoCA動物命名:} 貓 狗 老虎 豹，獅子，大笨象，牛，馬，狗，老虎，...\\
                \textbf{MoCA連續減法:} 一百減七啊，九啊七，九啊七啊，八八十三，八十三，再又...\\
                \textbf{MoCA數字廣度:} 二一 八 五四，四，四七二，\\
                \textbf{MoCA命名:} 獅子，啊，牛牛，駱駝，\\
                \textbf{MoCA句子重複:} 姨丈買魚腸，西施四十四歲，\\
                \textbf{MoCA抽象思維:} 大家 可以 交通用佢囉，即繫可以用可以，代代步啦，繫，...
                
                \vspace{0.1cm}
                請基於以上信息進行綜合評估。
                
                \vspace{0.2cm}
                \textbf{[Model Output]:} \\
                \texttt{\{\\
                \quad "cognitive\_assessment": \{\\
                \quad\quad "memory\_function": \{"status": "重度損害",\\
                \quad\quad\quad "evidence": ["HKLLT-4: 完全唔記得...", "HKLLT-5: 唔記得啦..."],\\
                \quad\quad\quad "interpretation": "嚴重記憶障礙，情景記憶功能嚴重受損"\},\\
                \quad\quad "executive\_function": \{"status": "損害", "interpretation": "抽象思維混亂"\},\\
                \quad\quad "attention\_working\_memory": \{"status": "損害", "interpretation": "計算錯誤頻繁"\},\\
                \quad\quad "language\_function": \{"status": "損害", "interpretation": "命名和重複錯誤"\}\\
                \quad \},\\
                \quad "step\_by\_step\_reasoning": \{\\
                \quad\quad "step\_1": \{"focus": "記憶功能分析", "conclusion": "記憶功能呈現重度損害"\},\\
                \quad\quad "step\_2": \{"focus": "多認知域分析", "conclusion": "多認知域損害模式"\},\\
                \quad\quad "step\_3": \{"focus": "風險整合", "conclusion": "存在較高認知下降風險"\}\\
                \quad \},\\
                \quad "diagnosis": \{"reasoning": "(1)記憶重度損害，(2)多認知域異常，(3)高風險因素",\\
                \quad\quad "confidence": 0.85, "is\_alzheimer\_likely": true\}\\
                \}}
                
            \end{minipage}%
            \hfill\vrule\hfill 
            \begin{minipage}[t]{0.48\linewidth}
                - Translated English Prompt -\\
                \textbf{[Instruction]:} \\
                \textbf{\#\# Clinical Protocol Context} \\
                ...
                
                \vspace{0.1cm}
                \textbf{\#\# Output Requirements}\\
                Please strictly output diagnostic results in the following JSON format:
                
                \vspace{0.1cm}
                \texttt{\{\\
                \quad "cognitive\_assessment": \{\\
                \quad\quad "memory\_function": \{"status": "normal/mild/moderate/severe impairment",\\
                \quad\quad\quad "evidence": ["HKLLT-4 performance: ...", "HKLLT-5 performance: ..."],\\
                \quad\quad\quad "interpretation": "brief explanation"\},\\
                \quad\quad "executive\_function": \{"status": "normal/impaired", ..."\},\\
                \quad\quad "attention\_working\_memory": \{"status": "normal/impaired", ..."\},\\
                \quad\quad "language\_function": \{"status": "normal/impaired", ..."\}\\
                \quad \},\\
                \quad "risk\_analysis": \{"main\_findings": [...], "risk\_factors": [...], "protective\_factors": [...]\},\\
                \quad "step\_by\_step\_reasoning": \{\\
                \quad\quad "step\_1": \{"focus": "memory function analysis", "observation": "...", "conclusion": "..."\},\\
                \quad\quad "step\_2": \{"focus": "multi-domain cognitive analysis", ..."\},\\
                \quad\quad "step\_3": \{"focus": "demographic factors \& risk integration", ..."\}\\
                \quad \},\\
                \quad "diagnosis": \{"reasoning": "...", "confidence": 0.0-1.0, "is\_alzheimer\_likely": true/false\}\\
                \}}
                
                \vspace{0.1cm}
                \textbf{Notes:}\\
                1. step\_by\_step\_reasoning must follow cognitive\_assessment results step-by-step\\
                2. diagnosis reasoning must cite conclusions from all 3 steps\\
                3. Remain objective; only judge as AD likely when multiple domains are impaired
                
                \vspace{0.1cm}
                \rule{\linewidth}{0.4pt}
                \vspace{0.1cm}
                
                \textbf{[User Input]:} \\
                \textbf{\#\# Current Case}\\
                \textbf{Subject Information}\\
                ID: TBS0134, Age: 70, Gender: Female, Education: 0.0 years
                
                \vspace{0.1cm}
                \textbf{Cognitive Test Transcripts}\\
                \textbf{HKLLT-4 (10-min delay):} Completely can''t remember, yes, completely forget, can''t recall any...\\
                \textbf{HKLLT-5 (30-min delay):} Ah, don''t remember, yes, yes...\\
                \textbf{MoCA Animal Naming:} Cat dog tiger leopard, lion, elephant, cow, horse, dog, tiger...\\
                \textbf{MoCA Serial 7s:} Hundred minus seven, ninety-seven, ninety-seven, eighty-three...\\
                \textbf{MoCA Digit Span:} Two one eight five four, four, four seven two...\\
                \textbf{MoCA Naming:} Lion, ah, cow cow, camel...\\
                \textbf{MoCA Sentence Repeat:} Uncle bought fish sausage, Xishi forty-four years old...\\
                \textbf{MoCA Abstraction:} Everyone can use for transport, can use, commute...
                
                \vspace{0.1cm}
                Please conduct comprehensive assessment based on above.
                
                \vspace{0.2cm}
                \textbf{[Model Output]:} \\
                \texttt{\{\\
                \quad "cognitive\_assessment": \{\\
                \quad\quad "memory\_function": \{"status": "severe impairment",\\
                \quad\quad\quad "evidence": ["HKLLT-4: Completely can''t remember...", "HKLLT-5: Don''t remember..."],\\
                \quad\quad\quad "interpretation": "Severe memory deficit, episodic memory severely impaired"\},\\
                \quad\quad "executive\_function": \{"status": "impaired", "interpretation": "Abstract thinking confused"\},\\
                \quad\quad "attention\_working\_memory": \{"status": "impaired", "interpretation": "Frequent calculation errors"\},\\
                \quad\quad "language\_function": \{"status": "impaired", "interpretation": "Naming and repetition errors"\}\\
                \quad \},\\
                \quad "step\_by\_step\_reasoning": \{\\
                \quad\quad "step\_1": \{"focus": "memory analysis", "conclusion": "Severe memory impairment"\},\\
                \quad\quad "step\_2": \{"focus": "multi-domain analysis", "conclusion": "Multi-domain damage pattern"\},\\
                \quad\quad "step\_3": \{"focus": "risk integration", "conclusion": "High cognitive decline risk"\}\\
                \quad \},\\
                \quad "diagnosis": \{"reasoning": "(1)Severe memory damage, (2)Multi-domain abnormal, (3)High risk",\\
                \quad\quad "confidence": 0.85, "is\_alzheimer\_likely": true\}\\
                \}}
                
            \end{minipage}
            
        \end{CJK*}
    \end{tcolorbox}}
    \caption{An example of the LLM-CoT baselineprompt with step-by-step reasoning for Alzheimer's diagnosis assessment. The left column shows the original Chinese prompt, and the right column shows the English translation.}
    \label{fig:llm-cot}
\end{figure*}
\section{Case Study}
\label{appendix:case}

We analyze representative error cases to characterize the failure modes of our framework, particularly where the verification loop proves insufficient.

\paragraph{Hallucination (Figure~\ref{fig:error1})} 
In the Serial 7 Subtraction task, the Examiner fabricates numbers not present in the transcript (e.g., 76, 69, 62). 
While the Verifier successfully detects these hallucinations and triggers re-generation, the Examiner responds by simply removing the flagged numbers rather than re-examining the transcript for valid responses. 
This reveals a limitation: the verification loop targets hallucination but cannot recover \textit{missing} valid answers that the Examiner failed to extract initially.

\paragraph{Boundary Case (Figure~\ref{fig:error2})} 
In the Abstraction task, the subject responds "use wheels to move" for the train-bicycle similarity question. 
The Examiner judges this as correct, but the Verifier overrides it as incorrect, reasoning that the response describes a concrete feature rather than an abstract category (transportation). 
However, this judgment conflicts with the human assessor's scoring, highlighting that boundary cases involving nuanced semantic distinctions remain challenging—even with verification,  and verifier judgments may not always align with human assessors on ambiguous cases.

\begin{figure*}[htbp] 
    \centering
    \scalebox{0.85}{
    \begin{tcolorbox}[colback=gray!5!white, colframe=gray!50!black, title=\textbf{Error Case 1: Hallucination (Bilingual Contrast)}]
        
        \begin{CJK*}{UTF8}{bsmi}
            \scriptsize 
            
            \begin{minipage}[t]{0.48\linewidth}
                - Original Prompt -\\
                \vspace{0.2cm}
                \textbf{[Instruction]:} \\
                \textbf{\#\# 任務介紹} \\
                你是一位醫學助理，評估MoCA連續減法任務。任務是從轉錄中提取受試者説出的所有有效數字回答。
                
                \vspace{0.1cm}
                \textbf{\#\# 注意事項}\\
                - 任務要求從100開始連續減7\\
                - 隻提取轉錄中明確出現的數字\\
                - 忽略重複、猶豫或修正的數字\\
                - 轉錄爲粵語口語\\
                - 不要捏造或推測未出現的數字

                \vspace{0.1cm}
                \textbf{\#\# 格式要求} \\
                有效回答: [數字1, 數字2, ...]\\
                \texttt{\textless tool\_call\textgreater}\\
                ...\\
                \texttt{\textless /tool\_call\textgreater}

                \vspace{0.1cm}
                \textbf{\#\# 範例} \\
                轉錄: \texttt{"九十三, 八十六, 七十九, 七十二"}\\
                有效回答: [93, 86, 79, 72]
                
                \vspace{0.1cm}
                \rule{\linewidth}{0.4pt} 
                \vspace{0.1cm}
                
                \textbf{[User Input]:} \\
                \textbf{\#\# 轉錄} \\
                請按格式要求輸出有效回答並調用函數。\\
                \\
                轉錄: \texttt{"由一 百減七, 百 一百開始囉喎, 一百減七 仲有九啊三 九啊三減七 就仲有, 九啊七減, 九啊七減 , 減個七咋喎繫咪啊, 一路減減個 即繫一路減七咋繫咪啊, 一百減七 仲有九啊三 九啊三減七就仲有, 啊 八啊 八啊四, 八啊四減七 仲有八啊一, 八啊一 減七就繫仲有六十, 六 十四 六十四減七就仲有, 五十, 嗯, 五十, 五十, 五十七, 五十七減 誒"}
                
                \vspace{0.2cm}
                \textbf{[Model Output (1st attempt):]} \\
                有效回答: [93, 84, 76, 69, 62]\\
                \texttt{\textless tool\_call\textgreater}\\
                ...\\
                \texttt{\textless /tool\_call\textgreater}
                
                \vspace{0.2cm}
                \textbf{[Verifier:]} \\
                數字 76 未在轉錄中找到，可能是錯誤識別或捏造。\\
                數字 69 未在轉錄中找到，可能是錯誤識別或捏造。\\
                數字 62 未在轉錄中找到，可能是錯誤識別或捏造。\\
                \\
                這次忽略之前的回答，重新思考。
                
                \vspace{0.2cm}
                \textbf{[Model Output (2nd attempt):]} \\
                有效回答: [93, 84, 79, 72, 65]\\
                \texttt{\textless tool\_call\textgreater}\\
                ...\\
                \texttt{\textless /tool\_call\textgreater}
                
                \vspace{0.2cm}
                \textbf{[Verifier:]} \\
                數字 79 未在轉錄中找到，可能是錯誤識別或捏造。\\
                數字 72 未在轉錄中找到，可能是錯誤識別或捏造。\\
                數字 65 未在轉錄中找到，可能是錯誤識別或捏造。\\
                \\
                這次忽略之前的回答，重新思考。
                
                \vspace{0.2cm}
                \textbf{[Final Model Output:]} \\
                有效回答: [93, 84]\\
                \texttt{\textless tool\_call\textgreater}\\
                ...\\
                \texttt{\textless /tool\_call\textgreater}

                \vspace{0.2cm}
                \textbf{[Verifier:]} \\
                pass, 所有數字都在轉錄中找到。
            \end{minipage}%
            \hfill\vrule\hfill 
            \begin{minipage}[t]{0.48\linewidth}
                - Translated English Prompt -\\

                \vspace{0.2cm}
                \textbf{[Instruction]:} \\
                \textbf{\#\# Task Description} \\
                You are a medical assistant evaluating the MoCA Serial Subtraction task. Extract all valid numerical answers spoken by the subject from the transcript.
                
                \vspace{0.1cm}
                \textbf{\#\# Notes}\\
                - Task requires subtracting 7 continuously from 100\\
                - Only extract numbers explicitly present in transcript\\
                - Ignore repeated, hesitant, or corrected numbers\\
                - Transcript is in Cantonese colloquial\\
                - Do not fabricate or infer numbers not present

                \vspace{0.1cm}
                \textbf{\#\# Format Requirements} \\
                Valid responses: [number1, number2, ...]\\
                \texttt{\textless tool\_call\textgreater}\\
                ...\\
                \texttt{\textless /tool\_call\textgreater}

                \vspace{0.1cm}
                \textbf{\#\# Example} \\
                Transcript: \texttt{"Ninety-three, eighty-six, seventy-nine, seventy-two"}\\
                Valid responses: [93, 86, 79, 72]
                
                \vspace{0.1cm}
                \rule{\linewidth}{0.4pt} 
                \vspace{0.1cm}
                
                \textbf{[User Input:]} \\
                \textbf{\#\# Transcript} \\
                Please output valid responses in required format and call function.\\
                \\
                Transcript: \texttt{"From one hundred minus seven, hundred start from hundred lo, hundred minus seven still have ninety-three, ninety-three minus seven still have, ninety-seven minus, ninety-seven minus, minus seven right, continuously minus means continuously minus seven right, hundred minus seven still have ninety-three, ninety-three minus seven still have, ah eighty ah eighty-four, eighty-four minus seven still have eighty-one, eighty-one minus seven is still have sixty, sixty-four, sixty-four minus seven still have, fifty, um, fifty, fifty, fifty-seven, fifty-seven minus eh"}
                
                \vspace{0.2cm}
                \textbf{[Model Output (1st attempt):]} \\
                Valid responses: [93, 84, 76, 69, 62]\\
                \texttt{\textless tool\_call\textgreater}\\
                ...\\
                \texttt{\textless /tool\_call\textgreater}
                
                \vspace{0.2cm}
                \textbf{[Verifier:]} \\
                Number 76 not found in transcript, may be misidentified or fabricated.\\
                Number 69 not found in transcript, may be misidentified or fabricated.\\
                Number 62 not found in transcript, may be misidentified or fabricated.\\
                \\
                Ignore previous answer and rethink.
                
                \vspace{0.2cm}
                \textbf{[Model Output (2nd attempt):]} \\
                Valid responses: [93, 84, 79, 72, 65]\\
                \texttt{\textless tool\_call\textgreater}\\
                ...\\
                \texttt{\textless /tool\_call\textgreater}
                
                \vspace{0.2cm}
                \textbf{[Verifier:]} \\
                Number 79 not found in transcript, may be misidentified or fabricated.\\
                Number 72 not found in transcript, may be misidentified or fabricated.\\
                Number 65 not found in transcript, may be misidentified or fabricated.\\
                \\
                Ignore previous answer and rethink.
                
                \vspace{0.2cm}
                \textbf{[Final Model Output:]} \\
                Valid responses: [93, 84]\\
                \texttt{\textless tool\_call\textgreater}\\
                ...\\
                \texttt{\textless /tool\_call\textgreater}

                \vspace{0.2cm}
                \textbf{[Verifier:]} \\
                Pass, all numbers found in transcript.
            \end{minipage}
            
        \end{CJK*}
    \end{tcolorbox}}
    \caption{
    An error case in the MoCA Serial Subtraction task.
    The examiner generates hallucinated numbers.
    With the verifier's feedback, the examiner simply removes hallucinated numbers and gives up on extracting other new numbers.
    The left column shows the original Cantonese prompt, and the right column shows the English translation.}
    \label{fig:error1}
\end{figure*}
\begin{figure*}[htbp] 
    \centering
    \scalebox{0.95}{
    \begin{tcolorbox}[colback=gray!5!white, colframe=gray!50!black, title=\textbf{Error Case 2: Boundary Case (Bilingual Contrast)}]
        
        \begin{CJK*}{UTF8}{bsmi}
            \scriptsize 
            
            \begin{minipage}[t]{0.48\linewidth}
                - Original Prompt -\\

                \vspace{0.2cm}
                \textbf{[Instruction]:} \\
                \textbf{\#\# 任務介紹} \\
                你是一位醫學助理，評估MoCA抽象概唸任務。任務是從轉錄中提取受試者對兩個問題的回答（Q1:火車-單車相似處, Q2:手表-直尺相似處），並判斷是否正確。
                
                \vspace{0.1cm}
                \textbf{\#\# 注意事項}\\
                - Q1和Q2之間有 \texttt{\textless|question-change|\textgreater} 分隔\\
                - Q1正確答案：交通工具、運輸工具、代步工具等（錯誤：有輪子、會動、沒區別、唔相似）\\
                - Q2正確答案：度量工具、測量工具、計量器具等（錯誤：有數字、可以看時間、不知道、諗唔到）\\
                - 轉錄爲粵語口語

                \vspace{0.1cm}
                \textbf{\#\# 格式要求} \\
                Q1回答: \textless 提取概唸\textgreater{} \quad Q1判斷: \texttt{\textless true/false\textgreater} \\
                Q2回答: \textless 提取概唸\textgreater{} \quad Q2判斷: \texttt{\textless true/false\textgreater} \\
                \texttt{\{ "Q1": \{"response": ["..."], "is\_correct": \textless true/false\textgreater\}, "Q2": \{...\} \}}

                \vspace{0.1cm}
                \textbf{\#\# 範例} \\
                轉錄: \texttt{"單車同火車, 都繫交通嘅工具, \textless|question-change|\textgreater, 手表, 間尺, 大家都繫, 計數目嘅..."} \\
                Q1回答: 交通工具 \quad Q1判斷: true \\
                Q2回答: 計數目 \quad \quad Q2判斷: false \\
                \texttt{\{ "Q1": \{..., "is\_correct": true\}, "Q2": \{..., "is\_correct": false\} \}}
                
                \vspace{0.1cm}
                \rule{\linewidth}{0.4pt} 
                \vspace{0.1cm}
                
                \textbf{[User Input]:} \\
                \textbf{\#\# 轉錄} \\
                請按格式要求輸出Q1、Q2回答和判斷，並輸出JSON結果。\\
                \\
                轉錄: \texttt{"嗯, 都繫用轆去行囉, 誒, 單車同火車, 用轆去行就可以繫, 噉我唔知啊, \textless|question-change|\textgreater, 手表同間尺相似地方喺邊度, 哦都繫長方形囉, 手表啊睇時間 間尺繫量度嘅, 哦都唔繫, 量度嘅嘢囉 我覺得就繫, 其實我真繫唔 都唔知有咩分別 有咩相同, 大家都繫長方形囉"}
                
                \vspace{0.2cm}
                \textbf{[Model Output (Intermediate)]:} \\
                Q1回答: 用轆去行 \quad Q1判斷: true \\
                Q2回答: 量度嘅嘢 \quad Q2判斷: true \\
                \texttt{\{ \\
                \quad "Q1": \{"response": ["用轆去行"], "is\_correct": true\}, \\
                \quad "Q2": \{"response": ["量度嘅嘢"], "is\_correct": true\} \\
                \}}
                
                \vspace{0.2cm}
                \textbf{[Verifier]:} \\
                Q1判斷錯誤（假陽性）：回答 "用轆去行" 不是抽象概唸，應該判爲錯誤 \\
                - 模型判斷: is\_correct = true \\
                - 正確判斷: is\_correct = false \\
                - 原因: 回答未包含交通工具/代步工具等抽象概唸，可能是具體特徵（如"有輪子"）\\
                修正建議：將 Q1.is\_correct 改爲 false
                
                \vspace{0.2cm}
                \textbf{[Final Model Output]:} \\
                \texttt{\{ \\
                \quad "Q1": \{"response": ["用轆去行"], "is\_correct": false\}, \\
                \quad "Q2": \{"response": ["量度嘅嘢"], "is\_correct": true\} \\
                \}}
            \end{minipage}%
            \hfill\vrule\hfill 
            \begin{minipage}[t]{0.48\linewidth}
                - Translated English Prompt -\\

                \vspace{0.2cm}
                \textbf{[Instruction]:} \\
                \textbf{\#\# Task Description} \\
                You are a medical assistant evaluating the MoCA Abstraction task. Extract the subject''s answers to two questions (Q1: Train-Bicycle similarity, Q2: Watch-Ruler similarity) from the transcript and determine correctness.
                
                \vspace{0.1cm}
                \textbf{\#\# Notes}\\
                - Q1 and Q2 are separated by \texttt{\textless|question-change|\textgreater}.\\
                - Q1 Correct: Transportation, Vehicle, Commute tools, etc. (Wrong: have wheels, move, no difference, not similar).\\
                - Q2 Correct: Measuring tools, Measurement instruments, Gauges, etc. (Wrong: have numbers, tells time, don''t know).\\
                - Transcript is in Cantonese colloquial.

                \vspace{0.1cm}
                \textbf{\#\# Format Requirements} \\
                Q1 Resp: \textless Concept\textgreater{} \quad Q1 Judge: \texttt{\textless true/false\textgreater} \\
                Q2 Resp: \textless Concept\textgreater{} \quad Q2 Judge: \texttt{\textless true/false\textgreater} \\
                \texttt{\{ "Q1": \{"response": ["..."], "is\_correct": \textless true/false\textgreater\}, "Q2": \{...\} \}}

                \vspace{0.1cm}
                \textbf{\#\# Example} \\
                Transcript: \texttt{"Bicycle and train, both are transport tools, \textless|question-change|\textgreater, watch, ruler, everyone is, counting numbers..."} \\
                Q1 Resp: Transport tools \quad Q1 Judge: true \\
                Q2 Resp: Counting numbers \quad Q2 Judge: false \\
                \texttt{\{ "Q1": \{..., "is\_correct": true\}, "Q2": \{..., "is\_correct": false\} \}}
                
                \vspace{0.1cm}
                \rule{\linewidth}{0.4pt} 
                \vspace{0.1cm}
                
                \textbf{[User Input]:} \\
                \textbf{\#\# Transcript} \\
                Please output Q1/Q2 response and judgment in JSON format.\\
                \\
                Transcript: \texttt{"Um, both use wheels to move, eh, bicycle and train, use wheels to move can be, then I don''t know, \textless|question-change|\textgreater, where''s the similarity between watch and ruler, oh both are rectangular, watch ah tells time ruler is for measuring, oh no, things for measuring lo I think is, actually I really don''t know what''s the difference what''s similar, both are rectangular"}
                
                \vspace{0.2cm}
                \textbf{[Model Output (Intermediate)]:} \\
                Q1 Resp: use wheels to move \quad Q1 Judge: true \\
                Q2 Resp: things for measuring \quad Q2 Judge: true \\
                \texttt{\{ \\
                \quad "Q1": \{"response": ["use wheels to move"], "is\_correct": true\}, \\
                \quad "Q2": \{"response": ["things for measuring"], "is\_correct": true\} \\
                \}}
                
                \vspace{0.2cm}
                \textbf{[Verifier]:} \\
                Q1 Judgment Error (False Positive): Response "use wheels to move" is not an abstract concept, should be incorrect \\
                - Model: is\_correct = true \\
                - Ground Truth: is\_correct = false \\
                - Reason: Response doesn''t include abstract concepts like transportation/vehicle, likely a concrete feature (e.g., "have wheels")\\
                Correction: Change Q1.is\_correct to false
                
                \vspace{0.2cm}
                \textbf{[Final Model Output]:} \\
                \texttt{\{ \\
                \quad "Q1": \{"response": ["use wheels to move"], "is\_correct": false\}, \\
                \quad "Q2": \{"response": ["things for measuring"], "is\_correct": true\} \\
                \}}
            \end{minipage}
            
        \end{CJK*}
    \end{tcolorbox}}
    \caption{An error case in the MoCA Abstraction task.
    In this boundary case, it''s hard to judge if "use wheels to move" is an abstract concept, while the assessor judges it as correct.
    The left column shows the original Cantonese prompt, and the right column shows the English translation.}
    \label{fig:error2}
\end{figure*}

\end{document}